\documentclass[times,8pt]{wlscirep}

\usepackage[utf8]{inputenc}
\usepackage[T1]{fontenc}
\usepackage{bm}
\usepackage{multicol}
\usepackage{float}

\usepackage[displaymath,mathlines]{lineno}
\title{ Measurement of the neutron charge radius and the role of its constituents}

\author[1]{H. Atac}
\author[1]{M. Constantinou}
\author[2,1]{Z.-E. Meziani}
\author[3]{M. Paolone}
\author[1,*]{N. Sparveris}
\affil[1]{Temple University, Philadelphia, PA 19122, USA}
\affil[2]{Argonne National Laboratory, Lemont, IL 60439, USA}
\affil[3]{New Mexico State University, Las Cruces, NM 88003, USA}

\affil[*]{corresponding author, e-mail: sparveri@temple.edu}

\begin{abstract}

The neutron is a cornerstone in our depiction of the visible universe. Despite the neutron zero-net electric charge, the asymmetric distribution of the positively- (up) and negatively-charged (down) quarks, a result of the complex quark-gluon dynamics, lead to a negative value for its squared charge radius, $\langle r_{\rm n}^2 \rangle$. The precise measurement of the neutron's charge radius thus emerges as an essential part of unraveling its structure. Here we report on a $\langle r_{\rm n}^2 \rangle$ measurement, based on the extraction of the neutron electric form factor, $G_{\rm E}^{\rm n}$, at low four-momentum transfer squared $(Q^2)$ by exploiting the long known connection between the $N \rightarrow \Delta$ quadrupole transitions and the neutron electric form factor. Our result, $\langle r_{\rm n}^2 \rangle = -0.110 \pm0.008~({\rm fm}^2)$, addresses long standing unresolved discrepancies in the $\langle r_{\rm n}^2 \rangle$ determination. The dynamics of the strong nuclear force can be viewed through the precise picture of the neutron's constituent distributions that result into the non-zero $\langle r_{\rm n}^2 \rangle$ value.

\end{abstract}

\begin{document}
\flushbottom
\maketitle
\thispagestyle{empty}
\begin{multicols}{2}

\quad\\
\textbf{Introduction} \\
The study of the nucleon charge radius has been historically instrumental towards the understanding of the nucleon structure. In the neutron case, it is the highly complicated dynamics of the strong force between quarks and gluons, the fermionic nature of quarks and spin-orbit correlations that leads to an asymmetric distribution of u- and d-quarks in it, thus resulting in a negative value for $\langle r_{\rm n}^2 \rangle$. The precise measurement of $\langle r_{\rm n}^2 \rangle$ becomes a critical part of our understanding of the nucleon dynamics. Furthermore, employing new, different techniques in extracting this fundamental quantity has proven most valuable, as recently exhibited in the proton's case: the disagreement of the proton charge radius, $r_{\rm p}$, as determined using the Lamb shift measurement in the muonic hydrogen atom~\cite{pohl:2010}, with the earlier results based on the hydrogen atom and the electron scattering measurement, gave rise to the proton radius puzzle~\cite{protonpuzz:2013}. In turn, this led to a significant reassessment of the methods and analyses utilized in the proton radius extraction, and to the consideration of physics beyond the standard model as potential solutions to this discrepancy. Various atomic and nuclear physics techniques were employed for the proton $r_{\rm p}$ measurement. However, in the neutron case, the determination of $\langle r_{\rm n}^2 \rangle$ is more challenging since no atomic method is possible and the electron scattering method suffers from severe limitations due to the absence of a free neutron target.
Thus, the extraction of $\langle r_{\rm n}^2 \rangle$ has been uniquely based on the measurement of the neutron-electron scattering length $b_{\rm ne}$, where low-energy neutrons are scattered by electrons bound in diamagnetic atoms. The $\langle r_{\rm n}^2 \rangle$ measurements adopted by the particle data group (PDG)~\cite{Kopecky:1997rw,Koester:1995nx,Aleksandrov:1986mw,Krohn:1973re} exhibit discrepancies, with the values ranging from $\langle r_{\rm n}^2 \rangle = -0.114 \pm0.003$~\cite{Koester:1995nx} to $\langle r_{\rm n}^2 \rangle = -0.134 \pm0.009~({\rm fm}^2)$~\cite{Aleksandrov:1986mw}. Among the plausible explanations that have been suggested for this, one can find the effect of resonance corrections and of the electric polarizability, as discussed e.g. in Ref.~\cite{Koester:1995nx}. However, these discrepancies have not been fully resolved, a direct indication of the limitations of this method.

An alternative way to determine $\langle r_{\rm n}^2 \rangle$ is offered by measuring the slope of the neutron electric form factor, $G_{\rm E}^{\rm n}$, at $Q^2 \rightarrow 0$, which is proportional to $\langle r_{\rm n}^2 \rangle$. In the past, determinations of $G_{\rm E}^{\rm n}$ at finite $Q^2$ were typically carried out by measuring double polarization observables in quasi-elastic electron scattering from polarized deuterium or $^3$He targets using polarized electron beams~\cite{Madey:2003av,Schlimme:2013eoz,Riordan:2010id,Glazier:2004ny,Plaster:2005cx,Zhu:2001md,Warren:2003ma,Rohe:1999sh,Passchier:1999ju,Bermuth:2003qh,Geis:2008aa,Eden.50.R1749,ostrick.83.276,golak.63.034006,Herberg:1999ud}. However, these measurements were not able to access $G_{\rm E}^{\rm n}$ at a sufficiently low $Q^2$ range so that the slope, and subsequently the $\langle r_{\rm n}^2 \rangle$ can be determined. 

In this work we rely on an alternative path to access $G_{\rm E}^{\rm n}$. It has long been known~\cite{Buchmann:2004ia,Vanderhaeghen:2007} that the ratios of the quadrupole to the magnetic dipole transition form factors (TFFs) of the proton, $C2/M1$ (CMR) and $E2/M1$ (EMR), are related to the neutron elastic form factors ratio $G_{\rm E}^{\rm n}/G_{\rm M}^{\rm n}$. Here, we follow that path and we access $G_{\rm E}^{\rm n}$ at low momentum transfers from high precision measurements of the two quadrupole TFFs. The main steps of this work are summarized here for clarity. First, we extract $G_{\rm E}^{\rm n}$ from the quadrupole TFF data, at low momentum transfers~\cite{Blomberg:2015zma,Stave:2006ea,Sparveris:2013ena,Sparveris:2006uk,Blomberg:2019caf}, utilizing the form factor relations~\cite{Buchmann:2004ia,Vanderhaeghen:2007} determined within the SU(6) and the large-$N_{\rm c}$ frameworks. The variance of the $G_{\rm E}^{\rm n}$ results from the two analyses is treated as a theoretical uncertainty. The $G_{\rm E}^{\rm n}(Q^2)$ form factor is then parametrized and fitted to the data, and $\langle r_{\rm n}^2 \rangle$ is determined from the $G_{\rm E}^{\rm n}$-slope at $Q^2=0$. Finally, we perform the flavor decomposition of the neutron and the proton form factors measurements and derive the flavor dependent quark densities in the nucleon, which reveal with high precision the role of the quark contributions to the neutron charge radius. 
\begin{figure}[H]
\centering
\includegraphics[width=\linewidth]{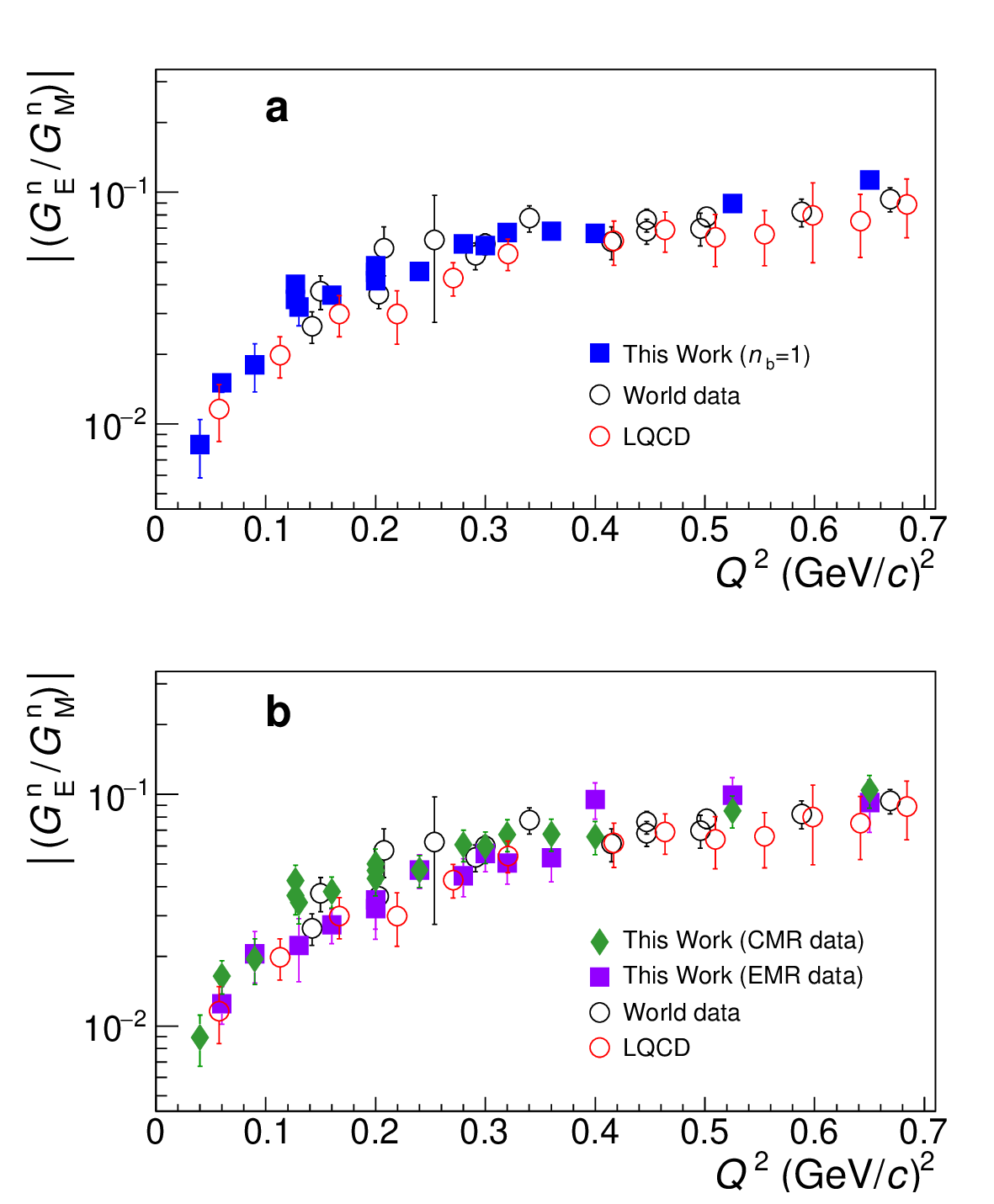}
\caption{{\bf The elastic neutron form factor ratio.} \\ {\bf a)} The neutron electric to magnetic form factor ratio $G_{\rm E}^{\rm n}/G_{\rm M}^{\rm n}$: world data~\cite{Madey:2003av,Schlimme:2013eoz,Riordan:2010id,Glazier:2004ny,Plaster:2005cx,Zhu:2001md,Warren:2003ma,Rohe:1999sh,Passchier:1999ju,Bermuth:2003qh,Geis:2008aa,Eden.50.R1749,ostrick.83.276,golak.63.034006,Herberg:1999ud} (black open circles), ratios calculated from the $N \rightarrow \Delta$ measurements~\cite{Blomberg:2015zma,Stave:2006ea,Sparveris:2013ena,Sparveris:2006uk,Sparveris:2004jn,Elsner:2005cz,Aznauryan:2009mx,Kelly:2005jy,Blomberg:2019caf} through Eq.~\ref{eqcmrgegmnb} for $n_{\rm b}=1$ (filled-squares), and lattice Quantum Chromodynamics (LQCD) results (red open circles)~\cite{Alexandrou:2018sjm}. \\ {\bf b)} The $G_{\rm E}^{\rm n}/G_{\rm M}^{\rm n}$ results from the large-$N_{\rm c}$ analysis of the Coulomb quadrupole measurements (CMR, filled diamonds) and of the Electric quadrupole measurements (EMR, filled boxes) from the experiments~\cite{Blomberg:2015zma,Stave:2006ea,Sparveris:2013ena,Sparveris:2006uk,Sparveris:2004jn,Elsner:2005cz,Aznauryan:2009mx,Kelly:2005jy,Blomberg:2019caf}. The neutron world data (black open circles) and the LQCD results (red open circles)~\cite{Alexandrou:2018sjm} are the same as in panel (a).\\
The error bars correspond to the total uncertainty, at the 1$\sigma$ or 68$\%$ confidence level.}
\label{fig-gegm}
\end{figure}

\quad\\
\textbf{Results} \\
A consequence of the SU(6) spin and flavor symmetry group which relates the nucleon and the $\Delta$ resonance leads to the following expression~\cite{Buchmann:2004ia}
\begin{linenomath}\begin{equation}\label{eqcmrgegmnb} \frac{G_{\rm E}^{\rm n}(Q^2)}{G_{\rm M}^{\rm n}(Q^2)}  =\frac{Q}{|\textbf{q}|} \frac{2Q}{M_{\rm N}} \frac{1}{n_{\rm b}(Q^2)} \frac{C2}{M1} (Q^2) \end{equation}\end{linenomath}
where $|\textbf{q}|$ is the virtual photon three-momentum transfer magnitude in the $\gamma$N center of mass frame and $M_{\rm N}$ is the nucleon mass. The $n_{\rm b}$ parametrizes the contribution from three-quark current terms, that tend to slightly increase the $C2/M1$ ratio (or correspondingly decrease the $G_{\rm E}^{\rm n}/G_{\rm M}^{\rm n}$), an SU(6) symmetry breaking correction that has been theoretically quantified to $\sim 10\%$~\cite{Buchmann:2004ia} (i.e. $n_{\rm b} \sim 1.1$). If one chooses to follow the most conservative path, a theoretical uncertainty can be assigned to this term that is equal to the full magnitude of the symmetry breaking contributions i.e. $n_{\rm b} = 1.1 \pm 0.1$. Considering the confidence with which the underlying theory is able to determine the level of the symmetry breaking contributions, the above assumption leads to a safe estimation, and most likely to an overestimation, of the theoretical uncertainty.

\begin{figure}[H]
\centering
\includegraphics[width=\linewidth]{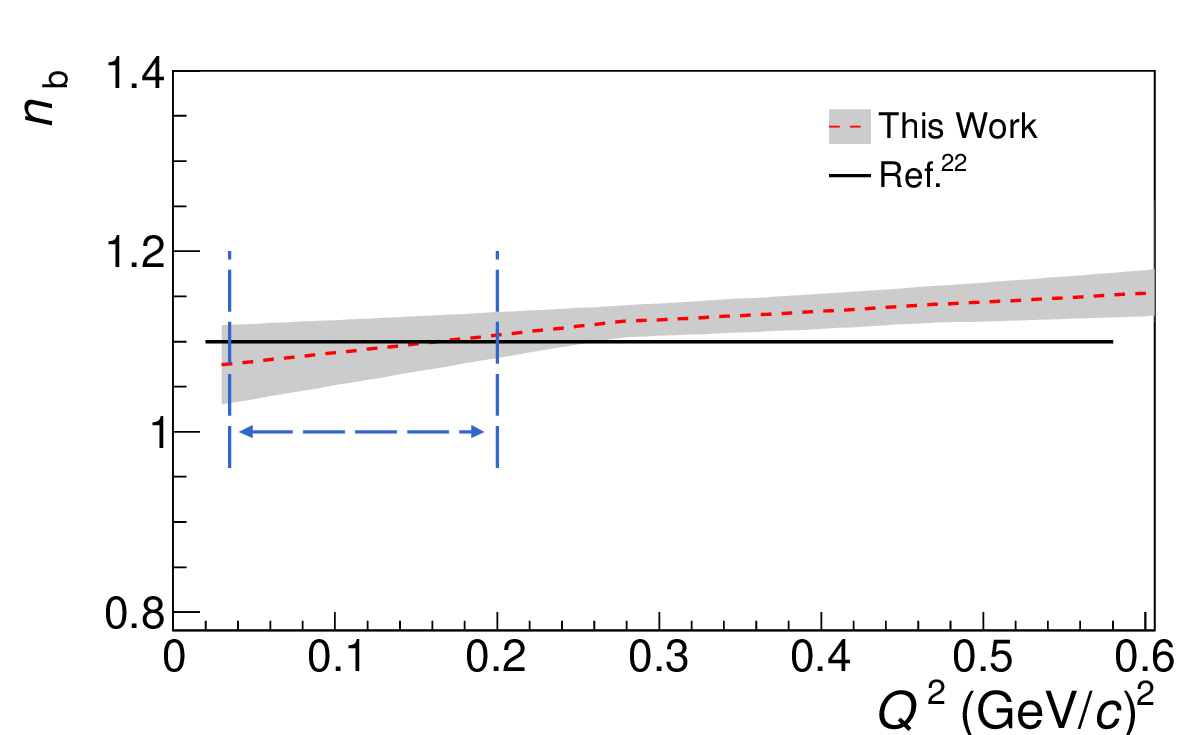}
\caption{{\bf The experimentally determined symmetry breaking contributions.} \\The breaking corrections $n_{\rm b}$ (dashed line) and the associated uncertainty $\delta n_{\rm b}$ (shaded band) at the 1$\sigma$ or 68$\%$ confidence level. The solid line indicates the $n_{\rm b}$ as theoretically determined in~\cite{Buchmann:2004ia}. The horizontal double-arrow marks the $Q^2$-range where the corrections have been employed for the measurement of $G_{\rm E}^{\rm n}$ in this work.}
\label{fig-nb}
\end{figure}

In one of the first steps of this work we check the validity of the underlying theory using experimental measurements. The wealth of the TFF~\cite{Blomberg:2015zma,Stave:2006ea,Sparveris:2013ena,Sparveris:2006uk,Sparveris:2004jn,Elsner:2005cz,Aznauryan:2009mx,Kelly:2005jy,Blomberg:2019caf} and of the $G_{\rm E}^{\rm n}/G_{\rm M}^{\rm n}$~\cite{Madey:2003av,Schlimme:2013eoz,Riordan:2010id,Glazier:2004ny,Plaster:2005cx,Zhu:2001md,Warren:2003ma,Rohe:1999sh,Passchier:1999ju,Bermuth:2003qh,Geis:2008aa,Eden.50.R1749,ostrick.83.276,golak.63.034006,Herberg:1999ud} world data allow to quantify the magnitude of the symmetry breaking corrections from the analysis of the experimental measurements. In Fig.~\ref{fig-gegm}a we show the neutron $G_{\rm E}^{\rm n}/G_{\rm M}^{\rm n}$ world data~\cite{Madey:2003av,Schlimme:2013eoz,Riordan:2010id,Glazier:2004ny,Plaster:2005cx,Zhu:2001md,Warren:2003ma,Rohe:1999sh,Bermuth:2003qh,Geis:2008aa,Eden.50.R1749,ostrick.83.276,golak.63.034006,Herberg:1999ud} (open black cirles), and we compare it to the $G_{\rm E}^{\rm n}/G_{\rm M}^{\rm n}$ ratios that we have derived from the TFFs $C2/M1$ measurements~\cite{Blomberg:2015zma,Stave:2006ea,Sparveris:2013ena,Sparveris:2006uk,Sparveris:2004jn,Elsner:2005cz,Aznauryan:2009mx,Kelly:2005jy,Blomberg:2019caf} (filled boxes) utilizing  Eq.~\ref{eqcmrgegmnb} with $n_{\rm b}=1$ (i.e. uncorrected for the symmetry breaking contributions). By parametrizing the two data sets and then forming their ratio we can experimentally determine the magnitude of the $n_{\rm b}(Q^2)$ contribution. A variety of functional forms have been explored to identify the functions that can provide a good fit to the data. All the appropriate functions that offer a good fit have been considered in the determination of $n_{\rm b}$ and the variance of the results arising from the choice of the functional form is adopted as an uncertainty. The procedure is further refined using lattice Quantum Chromodynamics (LQCD) results at low momentum transfers, where neutron data do not exist. In particular, we extracted the ratio $G_{\rm E}^{\rm n}/G_{\rm M}^{\rm n}$ from numerical simulations within LQCD using the $G_{\rm E}^{\rm n}$ and $G_{\rm M}^{\rm n}$ data of Ref.~\cite{Alexandrou:2018sjm}. The LQCD data provide further guidance on the $Q^2$-dependence of the $G_{\rm E}^{\rm n}/G_{\rm M}^{\rm n}$ ratio based on ab-initio QCD calculations, in a region where neutron form factor data are not available. The LQCD input results to a rather small refinement of $\leq 0.003$ in the determination of $n_{\rm b}$. The details on the determination of $n_{\rm b}(Q^2)$ are given in Section 2.1 of the Supplementary Information. The experimentally derived $n_{\rm b}(Q^2)$ is found in excellent agreement with the theoretical prediction~\cite{Buchmann:2004ia} as seen in Fig.~\ref{fig-nb}; this in-turn offers further credence to the theoretical effort in Ref.~\cite{Buchmann:2004ia}. Furthermore, the fitted parametrizations allow to constrain the $n_{\rm b}(Q^2)$ uncertainty by a factor of two compared to the most conservative $n_{\rm b} = 1.1 \pm 0.1$ (i.e. as indicated by the width of the uncertainty band in Fig.~\ref{fig-nb}), but also to determine these contributions accurately at very low momentum transfers where the analysis of the current TFF data takes place for the $G_{\rm E}^{\rm n}$ extraction.

The LQCD results entering our analysis are compared to the experimental world data and they exhibit a very good agreement as shown in Fig.~\ref{fig-gegm}a. The parameters of the LQCD calculation are such that they reproduce the physical value of the pion mass. Thus, such a calculation eliminates a major source of systematic uncertainties, that is, the need of a chiral extrapolation. Furthermore, the lattice results include both the connected and disconnected diagrams, and therefore $G_{\rm E}^{\rm n}$ and $G_{\rm M}^{\rm n}$ include both valence and sea quark contributions.

\begin{figure}[H]
\centering
\includegraphics[width=\linewidth]{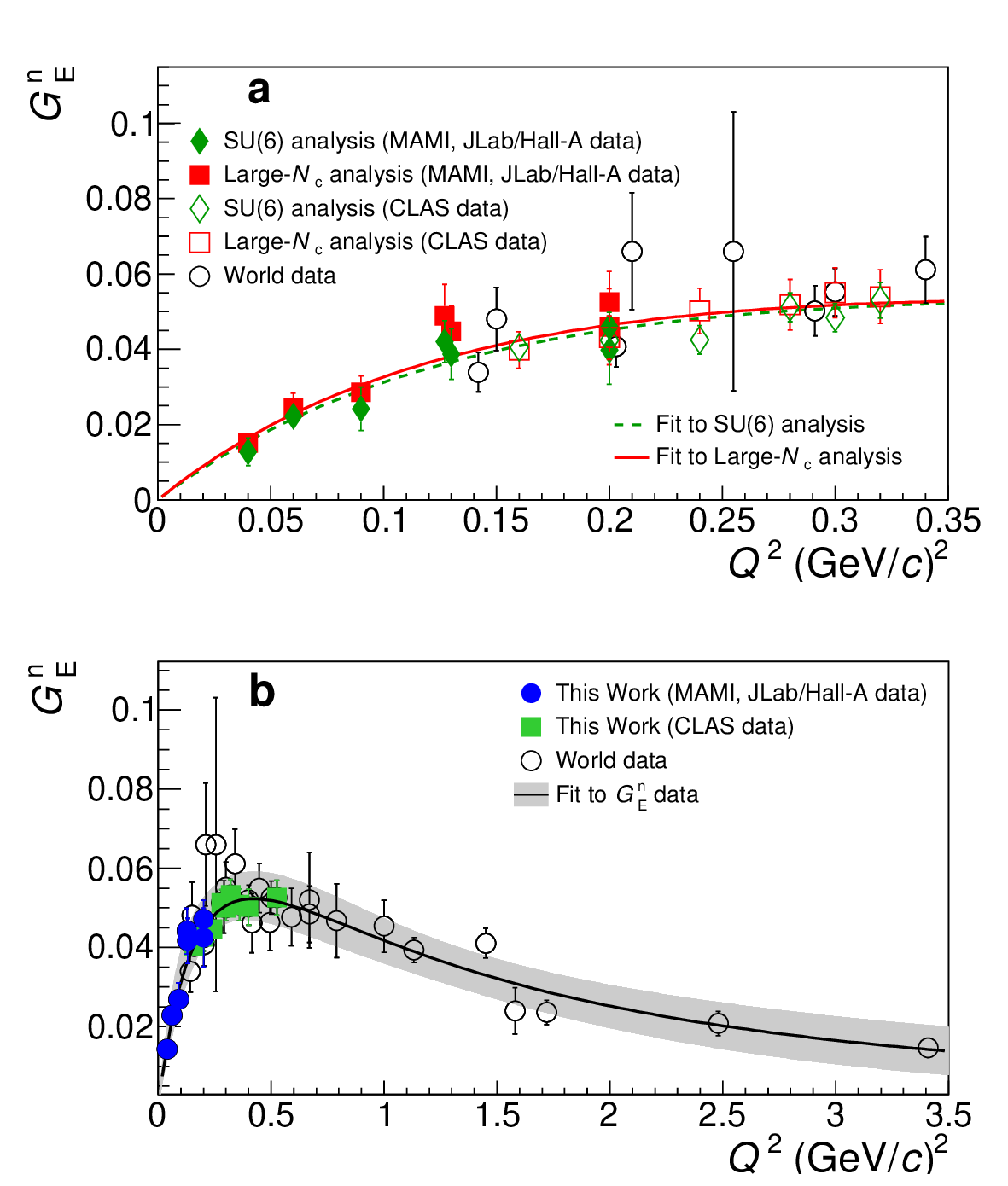}
\caption{{\bf The neutron electric form factor.} \\ {\bf a)} Green diamonds: the neutron electric form factor, $G_{\rm E}^{\rm n}$, at low momentum transfers from the analysis based on the SU(6)~\cite{Buchmann:2004ia} and $n_{\rm b}$ determined from the world data. Red boxes: the $G_{\rm E}^{\rm n}$ results from the analysis based on the large-$N_{\rm c}$~\cite{Vanderhaeghen:2007}. The fit to the data from the parametrization of Eq.~\ref{modgalster} is shown with the dashed and the solid curves, respectively. The filled symbols (diamonds/boxes) correspond to the analysis of the data from Ref.~\cite{Blomberg:2015zma,Stave:2006ea,Sparveris:2013ena,Sparveris:2006uk,Blomberg:2019caf} (MAMI, JLab/Hall-A data) and the open ones to that of Ref.~\cite{Aznauryan:2009mx} (CLAS data). \\
{\bf b)} Blue circles: The final $G_{\rm E}^{\rm n}$ results at low momentum transfers, extracted from the weighted average of the SU(6) and the large-$N_{\rm c}$ analysis results. The variance of the two data sets is quantified as a theoretical uncertainty. The solid curve shows the fit to the data from the parametrization of Eq.~\ref{modgalster}, with its uncertainty (shaded band). The $G_{\rm E}^{\rm n}$ world data (open-circles)~\cite{Madey:2003av,Schlimme:2013eoz,Riordan:2010id,Glazier:2004ny,Plaster:2005cx,Zhu:2001md,Warren:2003ma,Rohe:1999sh,Passchier:1999ju,Bermuth:2003qh,Geis:2008aa,Eden.50.R1749,ostrick.83.276,golak.63.034006,Herberg:1999ud} are shown. The extracted $G_{\rm E}^{\rm n}$ from the analysis of the CLAS measurements~\cite{Aznauryan:2009mx} at intermediate momentum transfers is also shown (green boxes).\\
The error bars correspond to the total uncertainty, at the 1$\sigma$ or 68$\%$ confidence level.}
\label{fig-gen}
\end{figure}

In our analysis we have extracted the $G_{\rm E}^{\rm n}$ under two scenarios; in one case we consider the conservative path where $n_{\rm b} = 1.1 \pm 0.1$, while in the second we utilize the $n_{\rm b}$ as we have determined it from the experimental world data. The two sets of results come to an agreement at the $\leq 3\%$ level; this is much smaller than to the overall $G_{\rm E}^{\rm n}$ uncertainty. A slightly improved $G_{\rm E}^{\rm n}$ uncertainty is obtained in the latter case due to an improved level of the $n_{\rm b}$ uncertainty, when these contributions are determined from the world data. The $G_{\rm E}^{\rm n}$ uncertainty of our results is driven by the following sources: i) Experimental (statistical and systematic) uncertainties in the determination of the $C2/M1$ ratio. ii) Uncertainties in the determination of $C2/M1$ due to the presence of non-resonant pion electro-production amplitudes that interfere with the extraction of the resonant amplitudes. These effects were studied by employing theoretical pion electro-production models in the data analysis; they were further investigated experimentally by measuring $C2/M1$ through an alternative reaction channel, the $p(e,e'p)\gamma$~\cite{Blomberg:2019caf}, where one employs a different theoretical framework for the ratio extraction (see Supplementary Information, Section 1). iii) The uncertainty of the symmetry breaking terms $\delta n_{\rm b}$, as discussed above.
iv) The uncertainty introduced by the choice of the $G_{\rm M}^{\rm n}$-parametrization, in order to extract the $G_{\rm E}^{\rm n}$ from the $G_{\rm E}^{\rm n}/G_{\rm M}^{\rm n}$ ratio (as typically done in such cases e.g.~\cite{Riordan:2010id,Geis:2008aa}). In this work we have used the one from Ref.~\cite{Ye:2017gyb} and we have quantified the associated uncertainty by repeating the analysis with alternative parametrizations. We have found a $\sim 0.5\%$ effect, which is rather small compared to the total $G_{\rm E}^{\rm n}$ uncertainty. The $G_{\rm E}^{\rm n}$ results are displayed in Fig.~\ref{fig-gen}a.

The relation between $G_{\rm E}^{\rm n}$ and the quadrupole transition form factors has also been established through large-$N_{\rm c}$ relations~\cite{Vanderhaeghen:2007}. The relations take the form
\begin{linenomath}\begin{equation}\label{marcemr}
\frac{E2}{M1} (Q^2) =\left (\frac{M_{\rm N}}{M_{\Delta}}\right)^{3/2} \frac{M_{\Delta}^2-M_{\rm N}^2}{2Q^2}\frac{G_{\rm E}^{\rm n}(Q^2)}{F_2^{\rm p}(Q^2)-F_2^{\rm n}(Q^2)}
 \end{equation}\end{linenomath}
 \begin{linenomath}\begin{equation}\label{marccmr}
\frac{C2}{M1} (Q^2) =\left (\frac{M_{\rm N}}{M_{\Delta}}\right)^{3/2} \frac{Q_+Q_-}{2Q^2}\frac{G_{\rm E}^{\rm n}(Q^2)}{F_2^{\rm p}(Q^2)-F_2^{\rm n}(Q^2)}
 \end{equation}\end{linenomath}
where $F_2^{\rm p(n)}$ are the nucleon Pauli form factors, $M_\Delta$ is the mass of the $\Delta$, and $Q_{\pm}=((M_{\Delta} \pm M_{\rm N})^2 + Q^2)^\frac{1}{2}$. Here one is free from any additional correction terms, such as the symmetry breaking contributions of Eq.~\ref{eqcmrgegmnb}. Another advantage is that the experimental database is extended to include the Electric quadrupole (E2) transition, which in turn allows for an improved extraction of $G_{\rm E}^{\rm n}$. Being able to extract $G_{\rm E}^{\rm n}$ independently through the Coulomb and the Electric quadrupole transitions offers a strong experimental test to the validity of the large-$N_{\rm c}$ relations and allows to quantify their level of theoretical uncertainty. The above relations come with a $15\%$ theoretical uncertainty~\cite{Vanderhaeghen:2007} that is treated accordingly in the $G_{\rm E}^{\rm n}$ analysis. The $G_{\rm E}^{\rm n}$ extraction from the Coulomb and from the Electric quadrupole transitions agree nicely within that level, as can be seen in Fig.~\ref{fig-gegm}b, and validate this level of uncertainty. For the well-known $G_{\rm E}^{\rm p}$, $G_{\rm M}^{\rm p}$ and $G_{\rm M}^{\rm n}$ that enter in the expressions through the Pauli form factors we have used recent parametrizations. For the $G_{\rm M}^{\rm p}$ and $G_{\rm M}^{\rm n}$ we used Ref.~\cite{Ye:2017gyb}. For $G_{\rm E}^{\rm p}$ we performed an updated parametrization  so that we may include recent measurements from Ref.~\cite{Xiong:2019umf} that were not yet available in Ref.~\cite{Ye:2017gyb} (see Supplementary Information, Section 4). 
For the large-$N_{\rm c}$ analysis the final results integrate both of the quadrupole transition form factors from each experiment, when both of them were simultaneously measured, into one $G_{\rm E}^{\rm n}$ measurement (see Supplementary Information, Section~2.2).
The extracted $G_{\rm E}^{\rm n}$ results from the large-$N_{\rm c}$ analysis are displayed in Fig.~\ref{fig-gen}a and are compared to the results from the SU(6) analysis in the same figure. The SU(6) analysis is in agreement with the large-$N_{\rm c}$ analysis $G_{\rm E}^{\rm n}$ results. For our final $G_{\rm E}^{\rm n}$ result we consider the weighted average of the two values, as shown in Fig.~\ref{fig-gen}b. The variance of the two values is treated as an additional $G_{\rm E}^{\rm n}$ theoretical uncertainty, and is accounted for accordingly in the $\langle r_{\rm n}^2 \rangle$ extraction.

\begin{figure}[H]
\centering
\includegraphics[width=\linewidth]{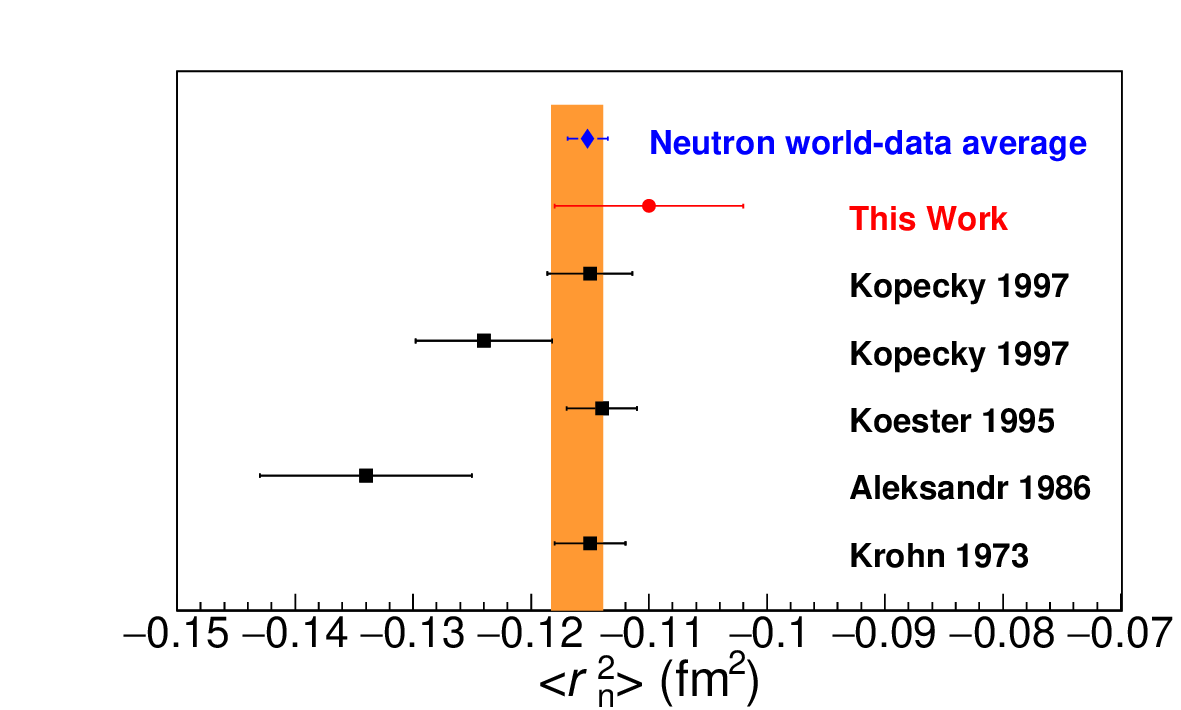}
\caption{{\bf The neutron mean square charge radius.} \\ The $\langle r_{\rm n}^2 \rangle$ measurement from this work
(red circle), with the error bar corresponding to the total uncertainty at the 1$\sigma$ or 68$\%$ confidence level, and from references~\cite{Kopecky:1997rw,Koester:1995nx,Aleksandrov:1986mw,Krohn:1973re} (black box) included in the PDG analysis for $\langle r_{\rm n}^2 \rangle$. The orange-band indicates the PDG averaged $\langle r_{\rm n}^2 \rangle$ value. The new weighted average of the world data is also shown (blue diamond) when the new $\langle r_{\rm n}^2 \rangle$ measurement reported in this work is included in the calculation.} 
\label{fig3}
\end{figure}

The neutron mean square charge radius is related to the slope of the neutron electric form factor as $Q^2 \rightarrow 0$ through 
\begin{linenomath}\begin{equation}\label{slope} \langle r_{\rm n}^2 \rangle = \left.-6 \frac{dG_{\rm E}^{\rm n}(Q^2)}{dQ^2} \right\rvert_{Q^2 \rightarrow 0}. \end{equation}\end{linenomath}
In order to determine the charge radius the data have to be fitted to a functional form, and the slope has to be determined at $Q^2=0$. It is important that a proper functional form is identified so that model dependent biases to the fit are avoided. In the past, the experimental data would allow to explore functional forms for $G_{\rm E}^{\rm n}(Q^2)$ with only two free parameters, that were lacking the ability to determine the neutron charge radius. The updated data allow to introduce an additional free parameter and to extract the $\langle r_{\rm n}^2 \rangle$ from measurements of the $G_{\rm E}^{\rm n}$ that was not possible previously. Our studies have shown that
\begin{linenomath}\begin{equation}\label{modgalster} G_{\rm E}^{\rm n}(Q^2)=(1 + Q^2/A)^{-2} \frac{B \tau}{1 + C \tau} , \end{equation}\end{linenomath}
is the most robust function for the radius extraction, where $\tau=Q^2/4M_{\rm N}^2$, and $A, B, C$ are free parameters (see Supplementary Information, Section~3). Our fits employ the $G_{\rm E}^{\rm n}$ data discussed in this work as well as the $G_{\rm E}^{\rm n}$ world data from~\cite{Madey:2003av,Schlimme:2013eoz,Riordan:2010id,Glazier:2004ny,Plaster:2005cx,Zhu:2001md,Warren:2003ma,Rohe:1999sh,Passchier:1999ju,Bermuth:2003qh,Geis:2008aa,Eden.50.R1749,ostrick.83.276,golak.63.034006,Herberg:1999ud}. 
The function describes the data very well, with a reduced $\chi^2$ of 0.74. The parameters obtained are $A = 0.505 \pm 0.079$~$({\rm GeV}/c)^2$, $B = 1.655 \pm 0.126$, $C = 0.909 \pm 0.583$, and $Q^2$ in units of $({\rm GeV}/c)^2$, leading to a value of  $\langle r_{\rm n}^2 \rangle = -0.110 \pm0.008~({\rm fm}^2)$, as shown in Fig.~\ref{fig3}. When the uncertainty of the symmetry breaking contributions in the SU(6) analysis is treated conservatively (i.e. $n_{\rm b}=1.1 \pm 0.1$) the final result becomes $\langle r_{\rm n}^2 \rangle = -0.109 \pm0.009~({\rm fm}^2)$ with a reduced $\chi^2$ of 0.74. 
Here we observe that the $\langle r_{\rm n}^2 \rangle$-uncertainty is not affected significantly by the different treatment of the symmetry breaking contributions in the two cases. 

The charge radius extraction is further explored through fits that are constrained within a limited range at low~$Q^2$ where $G_{\rm E}^{\rm n}$ remains monotonic, namely from $Q^2=$0 to $0.4~({\rm GeV}/c)^2$. In the fits to the data, the functional forms can be divided into two groups, those based on polynomials with varying orders and those that are based on rational forms (see Supplementary Information, Section 3.1). 
For the charge radius, the weighted average is extracted separately for each one of the two groups. A systematic uncertainty is also quantified within each group (i.e. a model uncertainty of the group) from the weighted variance of the results from all the fits within the group. The results from the two groups tend to have a similar overall uncertainty. A small systematic difference of the two group's central $\langle r_{\rm n}^2 \rangle$ values is observed, as studies over a varying fitting range have shown. For that reason a third uncertainty is determined: here we consider the spread of the two central values as indicative of the uncertainty that is associated with the choice of the group. Therefore, the final result is given by the average of the two group values for $\langle r_{\rm n}^2 \rangle$, while the half of the difference of the two values is assigned as an additional uncertainty. The details of the studies are presented in the Supplementary Information, Sections 3.1 and 3.2. The results from the low-$Q^2$ fits for all groups of functions and for variations to the fitting range are shown in the Supplementary Information, Tab.7 and Tab.8.
The low-$Q^2$ fits confirm the $\langle r_{\rm n}^2 \rangle$ extraction from the fits over the complete $G_{\rm E}^{\rm n}$ database, but they are vulnerable to model uncertainties that are associated with the choice of the fitted parametrization and they are not able to improve the $\langle r_{\rm n}^2 \rangle$ extraction. The studies indicate that more $G_{\rm E}^{\rm n}$ measurements are needed at lower momentum transfers so that a more competitive extraction can become possible from the low-$Q^2$ fits. This comes as no surprise when one considers the corresponding case for the proton, in which case the charge radius extraction from fits within a limited $Q^2$-range required measurements at significantly smaller momentum transfers, namely at $Q^2=0.0002~({\rm GeV}/c)^2$ - $0.06~({\rm GeV}/c)^2$~~\cite{Xiong:2019umf}.

The quadrupole TFF data at low momentum transfers~\cite{Blomberg:2015zma,Stave:2006ea,Sparveris:2013ena,Sparveris:2006uk,Blomberg:2019caf}, acquired at MAMI and at JLab/Hall-A, provide the critical $G_{\rm E}^{\rm n}$ data that were missing lower than $Q^2=0.20~({\rm GeV}/c)^2$ and make the $\langle r_{\rm n}^2 \rangle$ extraction possible. We now explore the potential of extending the current analysis to higher momentum transfers. This study aims to observe the effect of the $\langle r_{\rm n}^2 \rangle$ extraction if we enrich the $G_{\rm E}^{\rm n}$ database with measurements in the region where $G_{\rm E}^{\rm n}$ data already exist, higher than $Q^2=0.20~({\rm GeV}/c)^2$. The relations between the $G_{\rm E}^{\rm n}$ and the quadrupole transition form factors hold on very solid ground in the low~$Q^2$ region, i.e., lower than $Q^2=0.20~({\rm GeV}/c)^2$. On the other hand, they tend to hold less well at high momentum transfers. The relations do not come with a sharp $Q^2$ cut--off-value after which they do not hold, but one should avoid the $Q^2=1~({\rm GeV}/c)^2$ region where larger theoretical uncertainties could bias the charge radius extraction. In extending the database higher in $Q^2$, care has to be given so that any additional data at intermediate momentum transfers will benefit the fits without compromising the $\langle r_{\rm n}^2 \rangle$ extraction by the gradually increasing theoretical uncertainties. Our studies showed that when we integrate in the analysis the $G_{\rm E}^{\rm n}$ data that we extract from the CLAS measurements up to $Q^2=0.52~({\rm GeV}/c)^2$~~\cite{Aznauryan:2009mx} (see Fig.~\ref{fig-gen}b, green boxes) we find that $\langle r_{\rm n}^2 \rangle = -0.107 \pm0.007~({\rm fm}^2)$, compared to $\langle r_{\rm n}^2 \rangle = -0.110 \pm0.008~({\rm fm}^2)$ when these additional data are not included (see the Supplementary Information, Section 3.2 for details). Lastly, when the same data-set is included in the fits within the limited low-$Q^2$ range, as discussed in the previous paragraph, we find that $\langle r_{\rm n}^2 \rangle = -0.111 \pm0.006 \pm0.002_{\rm mod} \pm0.004_{\rm group}~({\rm fm}^2)$. Here the last two uncertainties (mod and group) are model-related uncertainties associated with the choice of the fitted parametrization (see Supplementary Information, Section 3.2 for details). We do not observe any additional benefit by extending the measurements higher in $Q^2$ since the fits' uncertainties do not improve when more data are included up to $Q^2=1~({\rm GeV}/c)^2$. In conclusion, a small benefit to the charge radius uncertainty can be observed when additional data up to $Q^2=0.5~({\rm GeV}/c)^2$ are utilized for the charge radius extraction. If one decides to eliminate any risk of introducing theoretical bias from the inclusion of the intermediate momentum transfer measurements for the final result, one can conservatively adopt the analysis that does not include these additional data, namely $\langle r_{\rm n}^2 \rangle = -0.110 \pm0.008~({\rm fm}^2)$.

\quad\\
\textbf{Discussion} \\
Our analysis and results offers valuable input towards addressing long standing unresolved $\langle r_{\rm n}^2 \rangle$ discrepancies of the $b_{\rm ne}$-measurements, which display a $\approx 10\%$  tension between the results, suggesting that there are still unidentified systematic uncertainties associated with this method of extraction. Our measurement is in disagreement with Ref.~\cite{Aleksandrov:1986mw} and supports the results of Ref.~\cite{Kopecky:1997rw,Koester:1995nx}. Considering that here we cross check $\langle r_{\rm n}^2 \rangle$ using a different extraction method, there is a strong argument so as to exclude the value of Ref.~\cite{Aleksandrov:1986mw} from the world data average. In such a case, the new weighted average value of the world data when we include our measurement and we exclude the one of Ref.~\cite{Aleksandrov:1986mw}, becomes $\langle r_{\rm n}^2 \rangle = -0.1152 \pm0.0017~({\rm fm}^2)$. Based on the current work, the particle data book value of $\langle r_{\rm n}^2 \rangle = -0.1161 \pm0.0022~({\rm fm}^2)$ is adjusted by $\sim 1\%$ and improves its uncertainty by $\sim 23\%$. We also note that our result agrees very well with a recent $\langle r_{\rm n}^2 \rangle$ calculation that is based on the determination of the deuteron structure radius in chiral effective field theory and utilizes atomic data for the difference of the deuteron and proton charge radii~\cite{bonnrn}.

\begin{figure}[H]
\centering
\includegraphics[width=\linewidth]{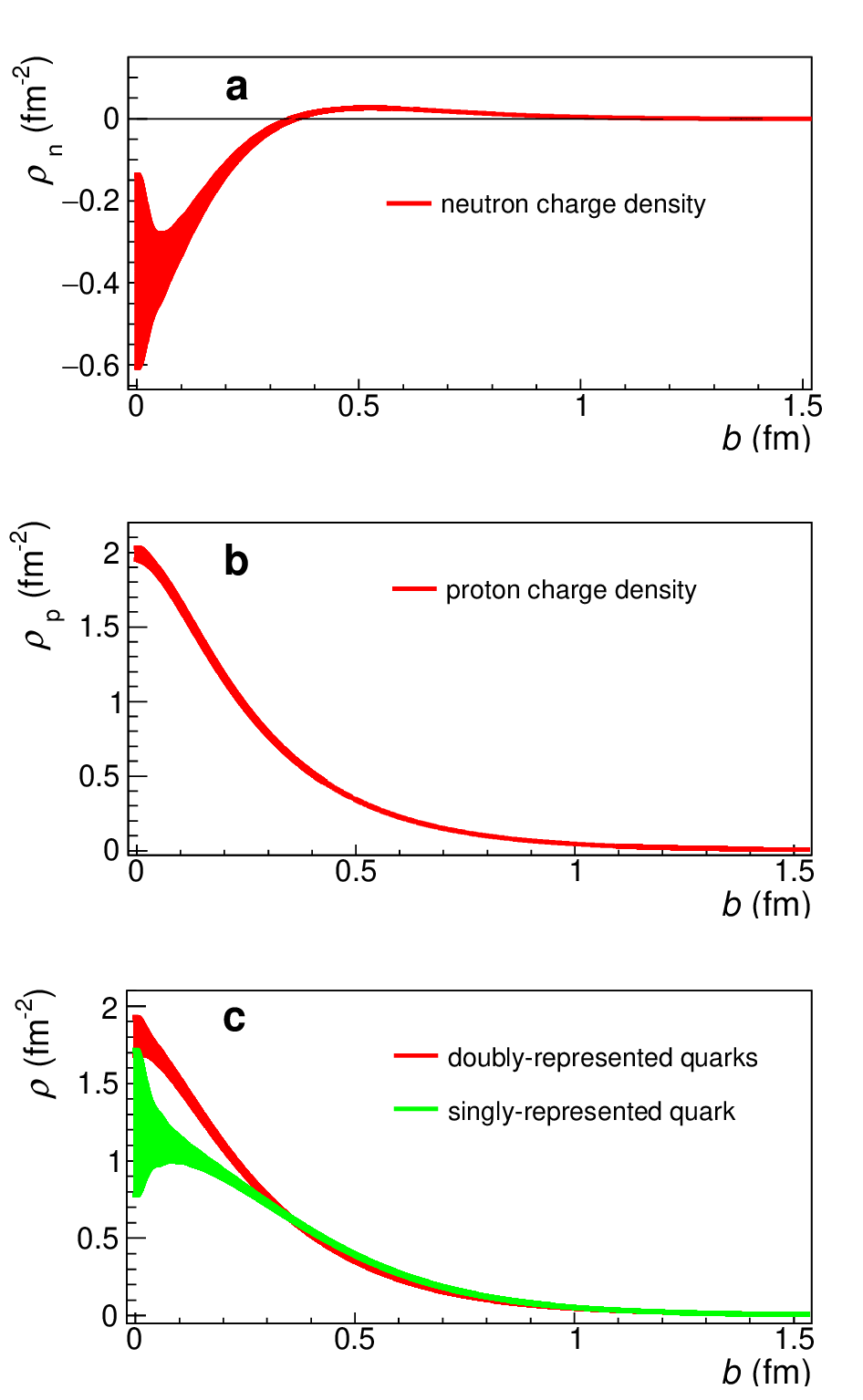}
\caption{{\bf The nucleon charge densities.}\\ {\bf a)} The neutron charge density $\rho_{\rm n}$. {\bf b)} The proton charge density $\rho_{\rm p}$. {\bf c)} Flavor decomposition of the nucleon charge densities: the doubly-represented (red) and singly-represented (green) charge densities in the nucleon. Each is normalized to unity.}
\label{fig-densall}
\end{figure}

The neutron's non-zero mean charge radius is a direct consequence of the asymmetric distribution of the positively-charged (up) and of the negatively-charged (down) quarks in the system, a consequence of the non-trivial quark gluon dynamics of the strong force. The quark distributions offer a detailed view as to how the non-zero $\langle r_{\rm n}^2 \rangle$ value arises. Here, one has to work on the infinite-momentum frame~\cite{Miller:2007uy} since it offers the inherent advantage that a true transverse charge density can be properly defined as the matrix element of a density operator between identical initial and final states. We find that the results of our analysis on $\langle r_{\rm n}^2 \rangle$ are particularly sensitive to the neutron's long-distance structure, and offer a significant improvement (factor of 2) in the precision of the neutron charge density at its surface (see Supplementary Information Fig.~9).
We extract the neutron and the proton charge densities at the infinite-momentum-frame from the most recent nucleon form factor parametrizations, where for $G_{\rm E}^{\rm n}$ we use the one determined in this work. The details are presented in the Supplementary Information, Section 4. The extracted neutron and proton charge densities are shown in Fig.~\ref{fig-densall}. From the two nucleon densities, invoking charge symmetry, and neglecting the $s\bar{s}$ contribution, we derive the $u$- and $d$-quark densities with an improved precision as shown in Fig.~\ref{fig-densall} (see Supplementary Information, Section~4 for details). The flavor dependent densities show that the singly-represented quark in the nucleon has a wider distribution compared to the doubly-represented quarks, which in turn exhibit a larger central quark density. Although the concentration of the two negatively charged quarks at the center of the neutron may appear to contradict the negative sign of the neutron's mean square charge radius, this is not truly the case. The 3D Breit frame and 2D infinite-momentum distributions are directly related to each other and the apparent discrepancies between the distributions in the two frames simply result from kinematical artifacts associated with spin~\cite{lorce}. The effect is rather dramatic in the neutron, where
the rest-frame magnetization is large and negative. The
contribution it induces competes with the convection contribution and gradually changes the sign at the center of
the charge distribution as one increases the momentum of the neutron. Thus, the appearance of a negative region around the center of the neutron
charge distribution in the infinite-momentum frame is
just a manifestation of the contribution induced by the
rest-frame magnetization.

In conclusion, we report on an alternative measurement of the neutron charge radius, based on the measurement of the neutron electric form factor $G_{\rm E}^{\rm n}$. An alternative path to the measurements based on the scattering of neutrons by electrons bound in diamagnetic atoms is presented. Our value of $\langle r_{\rm n}^2 \rangle = -0.110 \pm0.008~({\rm fm}^2)$ offers valuable input towards addressing long standing unresolved discrepancies in the $\langle r_{\rm n}^2 \rangle$ measurements, rejects earlier measurements, and improves the precision of the $\langle r_{\rm n}^2 \rangle$ world data average value. Furthermore, our data offer access to the associated dynamics of the strong nuclear force through the precise mapping of the quark distributions in the neutron that contribute to its non-zero charge radius. The current work lays the path for $\langle r_{\rm n}^2 \rangle$ measurements of higher precision. New experimental proposals based on this method , e.g. Jefferson Lab LOI 12-20-002, offer to improve the precision of the $\langle r_{\rm n}^2 \rangle$ measurement by nearly a factor of 2. Future experimental efforts will be able to utilize upgraded experimental setups that will fully exploit the advantages of this method. In particular, pushing the low momentum transfer limits of high precision measurements can lead to a further improvement in the precision of the $\langle r_{\rm n}^2 \rangle$ extraction.

\quad\\
\textbf{Methods} \\
Here we extract $G_{\rm E}^{\rm n}$, at low momentum transfers, from measurements of the quadrupole transition form factors. The neutron charge radius is then extracted from the slope of $G_{\rm E}^{\rm n}$ at $Q^2=0$. Utilizing the $G_{\rm E}^{\rm n}$ data and the world data for the nucleon elastic form factors the flavor decomposition of the nucleon electromagnetic form factors is performed, and the u- and d-quark distributions in the nucleon are extracted. The main steps of this work are as follows:
\begin{enumerate}

\item We extract $G_{\rm E}^{\rm n}$ from the Coulomb quadrupole and the Electric quadrupole transition form factor data at low momentum transfers~\cite{Blomberg:2015zma,Stave:2006ea,Sparveris:2013ena,Sparveris:2006uk} utilizing the form factor relations (Eqs.~\ref{eqcmrgegmnb},~\ref{marcemr}, and \ref{marccmr}) that have been determined within the SU(6)~\cite{Buchmann:2004ia} and the large-$N_{\rm c}$~\cite{Vanderhaeghen:2007} frameworks. The symmetry breaking corrections, $n_{\rm b}(Q^2)$, in Eq.~\ref{eqcmrgegmnb} are determined experimentally from the world data (elastic neutron form factors and $N\rightarrow\Delta$ transition form factors) and are further refined using state of the art Lattice QCD calculations. Parametrizations are used for the well known $G_{\rm E}^{\rm p}$, $G_{\rm M}^{\rm p}$ and $G_{\rm M}^{\rm n}$ form factors. Multiple parametrizations are employed so as to quantify the corresponding uncertainty introduced into the $G_{\rm E}^{\rm n}$ extraction.
\item The final $G_{\rm E}^{\rm n}$ values are extracted by the weighted average of the SU(6)~\cite{Buchmann:2004ia} and the large-$N_{\rm c}$~\cite{Vanderhaeghen:2007} analysis results. The variance between the results of the two methods is treated as a theoretical uncertainty.
\item The neutron mean square charge radius $\langle r_{\rm n}^2 \rangle$ is obtained from Eq.~\ref{slope} by fitting the $G_{\rm E}^{\rm n}$ data to the functional form in Eq.~\ref{modgalster} and determining the slope at $Q^2=0$. This functional form was shown to be the most robust function for the radius extraction from the neutron data. The fit employs additional $G_{\rm E}^{\rm n}$ data reported here and the $G_{\rm E}^{\rm n}$ world data extending to higher momentum transfers. 
\item The neutron and proton densities are derived at the infinite momentum frame through 
\begin{linenomath}\begin{equation}\label{eqrho}
\rho(b)= \int_0^\infty\ {dQ\ Q\ \over  2 \pi}J_0(Q b) {G_{\rm E}(Q^2)+\tau G_{\rm M}(Q^2)\over 1+\tau}
 \end{equation}\end{linenomath}
where $b$ is the transverse distance, $\tau={Q^2/4m^2}$ and $J_0$ the $0^{\rm th}$ order cylindrical Bessel function. Here we utilize the most recent parametrizations for the nucleon form factors, where for $G_{\rm E}^{\rm n}$ we use the one derived in this work. From the neutron and proton densities, invoking charge symmetry, and neglecting the $s\bar{s}$ contribution, we then extract the $u$- and $d$-quark densities in the proton (or doubly-represented and singly-represented quarks in the nucleon, respectively), where
\begin{linenomath}\begin{equation}\label{eqrhou}
\rho_{\rm u}(b)=\rho_{\rm p}(b)+\rho_{\rm n}(b)/2  
 \end{equation}\end{linenomath}
and
\begin{linenomath}\begin{equation}\label{eqrhod}
\rho_{\rm d}(b)=\rho_{\rm p}(b)+2\rho_{\rm n}(b).
\end{equation}\end{linenomath}
 
\end{enumerate}

\quad\\
\textbf{Data availability} \\
All the relevant data in this work are available from the authors upon request.
The data for the quadrupole TFFs used in this work are publicly available in their original publications~\cite{Blomberg:2015zma,Stave:2006ea,Sparveris:2013ena,Sparveris:2006uk,Blomberg:2019caf,Aznauryan:2009mx} where they are described in detail. The raw data from these experiments are archived in Jefferson Laboratory’s mass storage silo and at Temple University, Department of Physics. The $G_{\rm E}^{\rm n}$ data that we have used for the $\langle r_{\rm n}^2 \rangle$ extraction are available in the following sources: i) The new $G_{\rm E}^{\rm n}$ data that we have derived in this work, from the analysis of the quadrupole TFFs, are available in the Supplementary Information document. ii) The previously published $G_{\rm E}^{\rm n}$ world-data are available in their original publications~\cite{Madey:2003av,Schlimme:2013eoz,Riordan:2010id,Glazier:2004ny,Plaster:2005cx,Zhu:2001md,Warren:2003ma,Rohe:1999sh,Passchier:1999ju,Bermuth:2003qh,Geis:2008aa,Eden.50.R1749,ostrick.83.276,golak.63.034006,Herberg:1999ud}.

\quad\\
\textbf{Code availability} \\
The data analysis uses the standard C++ ROOT framework, which was developed at CERN and is freely available at https://root.cern.ch.
The computer codes used for the data analysis and for the generation of plots are available upon request. 



\section*{Acknowledgements}
We would like to thank M.~Vanderhaeghen as this work received great benefit from his input and suggestions. This work has been supported by the US Department of Energy Office of Science, office of Nuclear Physics under contract no. DE-SC0016577, DE-FG02-94ER40844 and DEAC02-06CH11357. M.C. acknowledges financial support by the U.S. Department of Energy, Office of Nuclear Physics, Early Career Award under Grant No.\ DE-SC0020405.

\section*{Author contributions}
N.S. and Z.-E.M. guided and supervised this effort. N.S. lead the experimental program at JLab (Hall A) and at MAMI (A1) that resulted to the new $G_{\rm E}^{\rm n}$ measurements (experiment spokesperson and analysis coordinator). H.A. and M.P. performed the fits, and the extraction of the charge densities and radii. M.C. was in charge of the LQCD results and supervised theoretical aspects of this work.

\section*{Competing interests}
The authors declare no competing interests. 

\section*{Supplementary information}
Supplementary information is available for this paper.

\section*{Correspondence and request of materials}
Correspondence and request of materials should be addressed to N.S.

\end{multicols}



\makeatother

\newpage
{\Large \title{Supplementary Information} }
\maketitle

{\Large

{\bf Measurement of the neutron charge radius and the role of its constituents}

\vspace{0.2cm}
H. Atac$^1$,
M. Constantinou$^1$,
Z.-E. Meziani$^{2,1}$,
M. Paolone$^3$,
and N. Sparveris$^{1,*}$ 
\vspace{0.5cm} \\
$^1$Temple University, Philadelphia, PA 19122, USA \\
$^2$Argonne National Laboratory, Lemont, IL 60439, USA \\
$^3$New Mexico State University, Las Cruces, NM 88003, USA \\
$^*$e-mail: sparveri@temple.edu \\

}
\setcounter{figure}{0}
\setcounter{equation}{0} 

\newpage

{\Large 
\centering {\bf Table of Contents}

\vspace{2cm}

\begin{table}[h]
\centering
\begin{tabular}{|clc|}
\hline
Section &  &  Page \\
\hline
               & & \\
1               & Description of the quadrupole transition   & 3       \\
               & form factor measurements    &        \\
               & & \\
2               & Extraction of $G_{\rm E}^{\rm n}$   & 3      \\
              & & \\
2.1               & Analysis within SU(6)   & 4      \\     
              & & \\
2.2               & Analysis within large-$N_{\rm c}$  & 7 \\    
              & & \\
3    & Neutron charge radius extraction   & 9      \\       
              & & \\   
3.1    & Extraction from fits within a limited low-$Q^2$ range   & 10      \\       
              & & \\   
3.2    & Extraction including additional data at higher $Q^2$   & 12      \\
              & & \\ 
3.2.1    & Considerations on the CLAS data uncertainties   & 14      \\       
              & & \\                 
4               & Flavor dependent charge densities   & 15       \\
               & & \\
5               & Lattice QCD results   & 15      \\
               & & \\
\hline
\end{tabular}
\end{table}

}

\newpage

In this document, we present details regarding to the analysis steps and the results reported in this work. More specifically we discuss: the experimental data of the quadrupole transition form factors and the nature of their uncertainties that enter into the $G_{\rm E}^{\rm n}$ extraction; the steps for the $G_{\rm E}^{\rm n}$ extraction from the transition form factors; the extraction procedure for $\langle r_{\rm n}^2 \rangle$; the extraction of the flavor dependent form factors, densities, and mean square radii; the Lattice QCD results. Detailed tables with all the parametrizations and the final results are given, as well as additional figures that are constructive towards the discussion of this work.

\section{Description of the quadrupole transition form factor measurements}

The neutron electric form factor $G_{\rm E}^{\rm n}$ has been extracted from measurements of the $N\rightarrow\Delta$ quadrupole transition form factors in the four-momentum transfer squared range $Q^2=0.04~({\rm GeV}/c)^2$ to $Q^2=0.20~({\rm GeV}/c)^2$.
The main aspects of these measurements that are relevant within the context of the current work are briefly described in this section, and for an extensive description of the experimental details we refer to~\cite{cBlomberg:2015zma,cStave:2006ea,cSparveris:2013ena,cSparveris:2006uk,cBlomberg:2019caf}. The measurements involve a coordinated experimental program, at JLab (Hall A) and at MAMI (A1), where:

\begin{itemize}
    \item The experiments shared the same methodology, experimental setup specifications, and analysis tools and procedures.
    \item The two experimental setups offer complementarity in their kinematical coverage of momentum transfer, and the measurements were coordinated accordingly so that the low-$Q^2$ region is optimally measured.
    \item Measurements at common kinematics have been performed (i.e. at $Q^2=0.13~({\rm GeV}/c)^2$ see Refs.~\cite{cSparveris:2013ena,cBlomberg:2015zma}) so that a cross check between the results from the two experimental setups is made, and consistency is ensured.
    \item A first measurement through the $p(e,e'p)\gamma$ channel~\cite{cBlomberg:2019caf} has been performed so that a comparison to the pion-electroproduction results~\cite{cBlomberg:2015zma,cStave:2006ea,cSparveris:2013ena,cSparveris:2006uk} can be made. Such a comparison offers critical input towards the model uncertainty of the world data. For optimal comparison, the measurements were performed at the same $Q^2$ and with the same experimental setup for both channels.  
\end{itemize}

More specifically, in both experiments an electron beam with energy $E \approx1$ GeV has been employed on a liquid hydrogen target, and two high resolution spectrometers are used to detect protons and electrons in coincidence for the measurement of the pion electro-production excitation channel. Both setups offer a spectrometer momentum resolution of $10^{-4}$, and a coincidence-time resolution between the proton and the electron spectrometers of $\approx1~ns$. The experimental setup offers high resolution i.e. the spectrometers focus within a very well defined bin of the kinematical phase space, and high precision cross section measurements are performed sequentially until the extended phase space has been covered (as opposed to different techniques where e.g. detectors with larger ($4\pi$) acceptance coverage but of reduced resolution are employed~\cite{cAznauryan:2009mx}). 

An important aspect of this experimental program is the consistent, and extensive, treatment of the non-resonant pion electro-production amplitudes that interfere with the extraction of the resonant amplitudes in the $N\rightarrow\Delta$ transition.
These interfering contributions, small in magnitude but large in number, can not be sufficiently constrained by the experimental measurements, and they thus result into a model uncertainty for the quadrupole transition form factors. In the past these contributions have been frequently poorly studied and rarely quoted as an uncertainty. Here, the effect of these amplitudes has been studied in the following manner:\\
(a) State of the art theoretical pion electroproduction models~\cite{cKamalov:1999hs, cSato:2000jf, cDrechsel:1998hk, cKamalov:2001qg, cArndt:2002xv} have been employed in the data analysis. The models offer different descriptions for the background amplitudes, leading to deviations in the extracted values of the transition form factors that are quantified as a model uncertainty.
This uncertainty is in turn appropriately treated in the extraction of the $G_{\rm E}^{\rm n}$. \\
(b) The above model uncertainties are determined within the pion electroproduction framework. However, the excitation can also be studied through the weak $p(e,e'p)\gamma$ channel. In this case the same physics signal can be extracted within a different theoretical framework, thus offering an ideal cross-check to the model uncertainties associated with the pion electroproduction channel. The branching ratio of the photon channel is very small ($0.6\%$), two orders of magnitude smaller compared to the pion-electroproduction, and as such it was not studied until recently. To that end, the first such measurement was conducted at MAMI (A1)~\cite{cBlomberg:2019caf}. Measurements at the same $Q^2$, utilizing the same experimental setup, were performed. The results were found in agreement between the two channels~\cite{cBlomberg:2019caf,cSparveris:2006uk}, thus giving credence to the quantification of the model uncertainties.

\section{Extraction of $G_{\rm E}^{\rm n}$}

The $G_{\rm E}^{\rm n}$ is extracted from the quadrupole transition form factor measurements~\cite{cBlomberg:2015zma,cStave:2006ea,cSparveris:2013ena,cSparveris:2006uk,cBlomberg:2019caf} utilizing the form factor relations~\cite{cBuchmann:2004ia,cVanderhaeghen:2007} determined within the SU(6) and the large-$N_{\rm c}$ frameworks. The data are analyzed independently within the two frameworks. The weighted average of the two values leads to the final $G_{\rm E}^{\rm n}$ and the variance of the two values is assigned as a $G_{\rm E}^{\rm n}$ theoretical uncertainty, that is accounted for accordingly in the $r_{\rm n}$ extraction. In order to extract $G_{\rm E}^{\rm n}$ from the $G_{\rm E}^{\rm n}/G_{\rm M}^{\rm n}$ we utilize a parametrization of the well known $G_{\rm M}^{\rm n}$ (as typically done in such cases, e.g.~\cite{cRiordan:2010id,cGeis:2008aa} etc). In this work we have used the recent parametrization~\cite{cYe:2017gyb}. The uncertainties associated with the $G_{\rm E}^{\rm n}$ extraction are the following:

\begin{itemize}
    \item Experimental (statistical and systematic) uncertainties of the quadrupole amplitudes.
    \item Model uncertainties of the quadrupole amplitudes (not applicable to measurement~\cite{cBlomberg:2019caf}).
    \item The theoretical uncertainty associated with the relations in~\cite{cBuchmann:2004ia,cVanderhaeghen:2007}.
    \item The uncertainty due to the $G_{\rm M}^{\rm n}$ parametrization.
\end{itemize}

The uncertainty introduced by $G_{\rm M}^{\rm n}$ has been studied by employing different $G_{\rm M}^{\rm n}$-parametrizations to the $G_{\rm E}^{\rm n}$ extraction. For that we have used the most recent~\cite{cYe:2017gyb}, the widely adopted in the past~\cite{cKelly:2004hm}, as well as a parametrization that we worked out towards that end. The overall effect to $G_{\rm E}^{\rm n}$ has been quantified $\approx0.5\%$, and is rather small compared to the total uncertainty.

\begin{table}[p]
\centering
\begin{tabular}{|l|l|l|l|l|l|l|l|}
\hline
       &\multicolumn{2}{c|}{$\sum\limits_{i=1}^{i} a_i x_i^i$}              & \multicolumn{4}{c|}{$\frac{\sum\limits_{i=1}^{i} b_i x_i^i}{(1 + \sum\limits_{j=1}^{j} c_{j} x_j^j)}$}  & \multicolumn{1}{c|}{$(1-\exp^{(d_1x)}$)}                \\ 
\hline
$F_{\rm R}(x)$ &\multicolumn{7}{c|}{ }                           \\ 
\hline
$a_1$ &1.420              & 1.670               & \quad -            & \quad -             & \quad -             &  \quad -             & \quad -              \\ 
$a_2$ &-2.351             & -5.962              & \quad -            & \quad -             & \quad -             &  \quad -             & \quad -              \\ 
$a_3$ &\quad -            & 10.968              & \quad -            & \quad -             & \quad -             &  \quad -             & \quad -              \\ 
$b_1$ &\quad -            & \quad -             & 1.679              &16.050               & 1.790               & 11.236               & \quad -              \\ 
$b_2$ &\quad -            & \quad -             & 14.840             &11.590               & 20.340              & 0.802                & \quad -              \\ 
$b_3$ &\quad -            & \quad -             &  \quad -           &15.168               & 0.801               & 15.168               & \quad -              \\ 
$b_4$ &\quad -            & \quad -             &  \quad -           & \quad -             & \quad -             & 43.971               & \quad -              \\ 
$c_1$ &\quad -            & \quad -             & 13.602             &43.980               & 18.111              & 1.756                & \quad -              \\ 
$c_2$ &\quad -            & \quad -             & 31.396             &1.757                & 49.377              & 16.089               & \quad -              \\ 
$c_3$ &\quad -            & \quad -             & \quad -            & \quad -             & -40.446             &  \quad -             & \quad -              \\ 
$d_1$ &\quad -            & \quad -             & \quad -            & \quad -             & \quad -             &  \quad -             & -0.336               \\ 
\hline
$F^*_{\rm R}(x)$ &\multicolumn{7}{c|}{ }                           \\ 
\hline
$a_1$ & 1.587             &1.707            & \quad -         & \quad -          & \quad -         & \quad -          & \quad -            \\ 
$a_2$ & -2.243            &-4.349           & \quad -         & \quad -          & \quad -         & \quad -          & \quad -            \\ 
$a_3$ & \quad -           &6.078            & \quad -         & \quad -          & \quad -         & \quad -          & \quad -            \\ 
$b_1$ & \quad -           &\quad -          & 1.851           &0.170             &1.919            & 12.346           & \quad -            \\ 
$b_2$ & \quad -           &\quad -          & 2.495           &15.730            &23.081           & 0.800            & \quad -            \\ 
$b_3$ & \quad -           &\quad -          &  \quad -        &3.455             &0.801            & 15.369           & \quad -            \\ 
$b_4$ &\quad -            &\quad -          &  \quad -        & \quad -          &\quad -          & 35.134           & \quad -            \\ 
$c_1$ &\quad -            &\quad -          & 6.090           & 15.298           &18.943           & 1.904            & \quad -            \\ 
$c_2$ &\quad -            &\quad -          & -0.955          & 1.793            &34.286           & 17.579           & \quad -            \\ 
$c_3$ &\quad -            &\quad -          & \quad -         & \quad -          &-12.404          & \quad -          & \quad -            \\ 
$d_1$ &\quad -            &\quad -          & \quad -         & \quad -          & \quad -         & \quad -          & -0.379             \\ 
\hline
$F^{\rm l}_{\rm R}(x)$ &\multicolumn{7}{c|}{ }                           \\ 
\hline
$a_1$ & 1.240             &1.332            & \quad -         & \quad -          & \quad -            & \quad -            & \quad -            \\ 
$a_2$ & -2.102            &-3.833           & \quad -         & \quad -          & \quad -            & \quad -            & \quad -            \\ 
$a_3$ & \quad -           &6.228            & \quad -         & \quad -          & \quad -            & \quad -            & \quad -            \\ 
$b_1$ & \quad -           &\quad -          & 1.363           & 1.774            & 0.996              & 9.800              & \quad -            \\ 
$b_2$ & \quad -           &\quad -          & 2.816           & 11.497           & 27.413             & 0.801              & \quad -            \\ 
$b_3$ & \quad -           &\quad -          & \quad -         & 3.934            & 0.802              & 15.753             & \quad -            \\ 
$b_4$ &\quad -            &\quad -          & \quad -         & \quad -          & \quad -            & 48.822             & \quad -            \\ 
$c_1$ &\quad -            &\quad -          & 5.882           & 20.740           & 16.773             & 1.366              & \quad -            \\ 
$c_2$ &\quad -            &\quad -          & 4.267           & 1.326            & 58.324             & 16.228             & \quad -            \\ 
$c_3$ &\quad -            &\quad -          & \quad -         & \quad -          & 254.194            & \quad -            & \quad -            \\ 
$d_1$ &\quad -            &\quad -          & \quad -         & \quad -          & \quad -            & \quad -            & -0.296             \\

\hline

\end{tabular}
\caption{\label{nbdat}  The fitted parameters of the $F_{\rm R}(x)$, $F^*_{\rm R}(x)$, and $F^{\rm l}_{\rm R}(x)$ functions.}

\end{table}

\subsection{Analysis within SU(6)}

A consequence of the SU(6) spin and flavor symmetry group in which the nucleon and the $\Delta$ resonance belong leads to the following expression~\cite{cBuchmann:2004ia}
\begin{linenomath}\begin{equation}\label{eqcmrgegmnb} \frac{G_{\rm E}^{\rm n}(Q^2)}{G_{\rm M}^{\rm n}(Q^2)}  =\frac{Q}{|\textbf{q}|} \frac{2Q}{M_{\rm N}} \frac{1}{n_{\rm b}(Q^2)} \frac{C2}{M1} (Q^2) \end{equation}\end{linenomath}
where $|\textbf{q}|$ is the virtual photon three-momentum transfer magnitude in the $\gamma$N center of mass frame, $M_{\rm N}$ is the nucleon mass, and $n_{\rm b}$ describes three-quark current terms that slightly increase the C2/M1 ratio (or correspondingly decrease the $G_{\rm E}^{\rm n}/G_{\rm M}^{\rm n}$), an SU(6) symmetry breaking correction that has been theoretically quantified to $\approx 10\%$~\cite{cBuchmann:2004ia} (i.e. $n_{\rm b} \approx 1.1$). \\

The data have been analyzed in two ways:\\

i) Following the most conservative path, a theoretical uncertainty can be assigned that is equal to the full magnitude of the symmetry breaking contributions i.e. $n_{\rm b} = 1.1 \pm 0.1$. Considering the confidence with which the underlying theory is able to determine the level of the symmetry breaking contributions, the above assumption leads to a safe estimation, and most likely to an overestimation, of the theoretical uncertainty. \\

ii) The wealth of the experimental world data for $C2/M1$~\cite{cBlomberg:2015zma,cStave:2006ea,cSparveris:2013ena,cSparveris:2006uk,cSparveris:2004jn,cElsner:2005cz,cAznauryan:2009mx,cKelly:2005jy,cBlomberg:2019caf} and for $G_{\rm E}^{\rm n}/G_{\rm M}^{\rm n}$~\cite{cMadey:2003av,cSchlimme:2013eoz,cRiordan:2010id,cGlazier:2004ny,cPlaster:2005cx,cZhu:2001md,cWarren:2003ma,cRohe:1999sh,cBermuth:2003qh,cGeis:2008aa,cEden.50.R1749,costrick.83.276,cgolak.63.034006,cHerberg:1999ud} allow to determine the magnitude of the symmetry breaking corrections \footnote{A revised analysis of the~\cite{cStave:2006ea} has updated the C2/M1 result to $(-4.10 \pm0.27_{\rm stat+sys} \pm0.26_{\rm mod})\%$ }.  Refinements to the corrections are implemented utilizing the LQCD data for $G_{\rm E}^{\rm n}/G_{\rm M}^{\rm n}$ reported in this work (see later in the LQCD section). \\

The results from (i) and (ii) are in excellent agreement; the difference between the analysis (i) and (ii) results to a very small effect in the $r_{\rm n}$ extraction (see the radius extraction section). This comes as a consequence of the confirmation of the $n_{\rm b}$ theoretical prediction~\cite{cBuchmann:2004ia} by the experimental data (see Fig.~\ref{fig-nb}). The experimental determination of the breaking corrections $n_{\rm b}$ takes place as follows: \\

First, we determine the set of appropriate functions $F_{\rm R}(Q^2)$ that can successfully parametrize the $G_{\rm E}^{\rm n}/G_{\rm M}^{\rm n}$ ratio.
To that end we have identified the following forms that are able to provide a good fit to the data: \\

\qquad $F_{\rm R}(x) = \sum\limits_{i=1}^{j} a_i x_i^i$, \quad
$F_{\rm R}(x) =\frac{\sum\limits_{i=1}^{j} b_i x_i^i}{(1 + \sum\limits_{j=1}^{j} c_{j} x_j^j)}$, \quad
$F_{\rm R}(x) =  (1-\exp^{(d_1x)})$. \\

For each one of these functions: 

\begin{itemize}
    \item The $F_{\rm R}(Q^2)$ parametrization is determined by fitting to the $G_{\rm E}^{\rm n}/G_{\rm M}^{\rm n}$ world data~\cite{cMadey:2003av,cSchlimme:2013eoz,cRiordan:2010id,cGlazier:2004ny,cPlaster:2005cx,cZhu:2001md,cWarren:2003ma,cRohe:1999sh,cBermuth:2003qh,cGeis:2008aa,cEden.50.R1749,costrick.83.276,cgolak.63.034006,cHerberg:1999ud}.
    
    \item The $F^*_{\rm R}(Q^2)$ is determined by fitting to the $G_{\rm E}^{\rm n}/G_{\rm M}^{\rm n}$ ratios as derived from the $N\rightarrow\Delta$ measurements~\cite{cBlomberg:2015zma,cStave:2006ea,cSparveris:2013ena,cSparveris:2006uk,cSparveris:2004jn,cElsner:2005cz,cAznauryan:2009mx,cKelly:2005jy,cBlomberg:2019caf} through Eq.~\ref{eqcmrgegmnb} for $n_{\rm b}$=1.
    
    \item The breaking corrections are then determined, for each functional form, through: $n_{\rm b}(Q^2)=F^*_{\rm R}(Q^2)/F_{\rm R}(Q^2)$.
    
    \item The procedure is then repeated for all appropriate functions.
    
    \item We combine all of the above $n_{\rm b}(Q^2)$ results (i.e. as derived for all the functional forms) and we experimentally determine the $n_{\rm b}(Q^2) \pm \delta n_{\rm b}(Q^2)$: The $n_{\rm b}(Q^2) = [n_{\rm b(max)}(Q^2)+n_{\rm b(min)}(Q^2)]/2$, where $n_{\rm b(max)}$ and $n_{\rm b(min)}$ are the maximum and minimum $n_{\rm b}$ derived at a given $Q^2$. The $\delta n_{\rm b}(Q^2)$ results from the maximum spread of the $n_{\rm b}(Q^2)$ solutions at any given $Q^2$, i.e. $[n_{\rm b(max)}(Q^2)-n_{\rm b(min)}(Q^2)]$.
    
\end{itemize}

Since the neutron data do not extend lower than $Q^2=0.14~({\rm GeV}/c)^2$
we have decided to explore refinements in the determination of $n_{\rm b}$ by utilizing state of the art LQCD calculations for the $G_{\rm E}^{\rm n}/G_{\rm M}^{\rm n}$ ratio that extend lower in $Q^2$. The LQCD results bring valuable input in describing the $Q^2$-dependence of $G_{\rm E}^{\rm n}/G_{\rm M}^{\rm n}$ in this region based on ab-initio QCD calculations, thus having the potential to drive more accurately the fits. The procedure followed here is the following:

\begin{itemize}

    \item First, the LQCD $G_{\rm E}^{\rm n}/G_{\rm M}^{\rm n}$ results are normalized to the neutron world data, within the region where there is overlap for both data sets i.e. $Q^2 \ge 0.14~({\rm GeV}/c)^2$. The LQCD results agree remarkably well to the experimental data but we nevertheless introduce this normalization in order to absolutely baseline the two data sets within the $Q^2$ range that they overlap.
    
    \item The $F^{\rm l}_{\rm R}(k)$ is then determined by fitting the LQCD data to the same set of functions, as in the previous step, for the full momentum transfer range of the LQCD data-set so that the data at $Q^2 < 0.14~({\rm GeV}/c)^2$ are included.
    
    \item The breaking corrections are now given by: $n_{\rm b}(k)=F^*_{\rm R}(k)/F^{\rm l}_{\rm R}(k)$
    
\end{itemize}

For $Q^2 \ge 0.14~({\rm GeV}/c)^2$ the corrections are naturally identical to the ones determined using the experimental neutron data, since there the LQCD ratios have been normalized to the experimental data.
For the lower $Q^2$ region, the resulting refinement to the $n_{\rm b}$ determination is $\leq \pm0.3\%$ i.e. $\delta n_{\rm b}(Q^2)\leq \pm0.003$. That is an order of magnitude smaller than the total experimental $\delta n_{\rm b}$ uncertainty, that ranges between $\pm2.5\%$ to $\pm4\%$ in that region. Furthermore, we do not allow these refinements to reduce the $\delta n_{\rm b}$ uncertainty in any way, but only to increase it where the results extend it further than what was determined using the experimental data only. The fitted parameters for the $F_{\rm R}(x)$, $F^*_{\rm R}(x)$, and $F^{\rm l}_{\rm R}(x)$ are given in Table~\ref{nbdat}. The experimentally determined $n_{\rm b}(Q^2)$ is shown in Fig.~\ref{fig-nb} and the shaded band corresponds to the $n_{\rm b}$ uncertainty. The solid line marks the $n_{\rm b}$ as theoretically determined in~\cite{cBuchmann:2004ia}, and one can note the remarkable agreement between the theoretical and the experimental values. The $n_{\rm b}$ results are also listed in Table~\ref{tabnbc} for the $Q^2$ range that they have been used for the $G_{\rm E}^{\rm n}$ extraction.
The $G_{\rm E}^{\rm n}$ results and the breakdown of the uncertainties are given in Table~\ref{newtable}.

\begin{figure}[t]
\centering
\includegraphics[width=11.0cm]{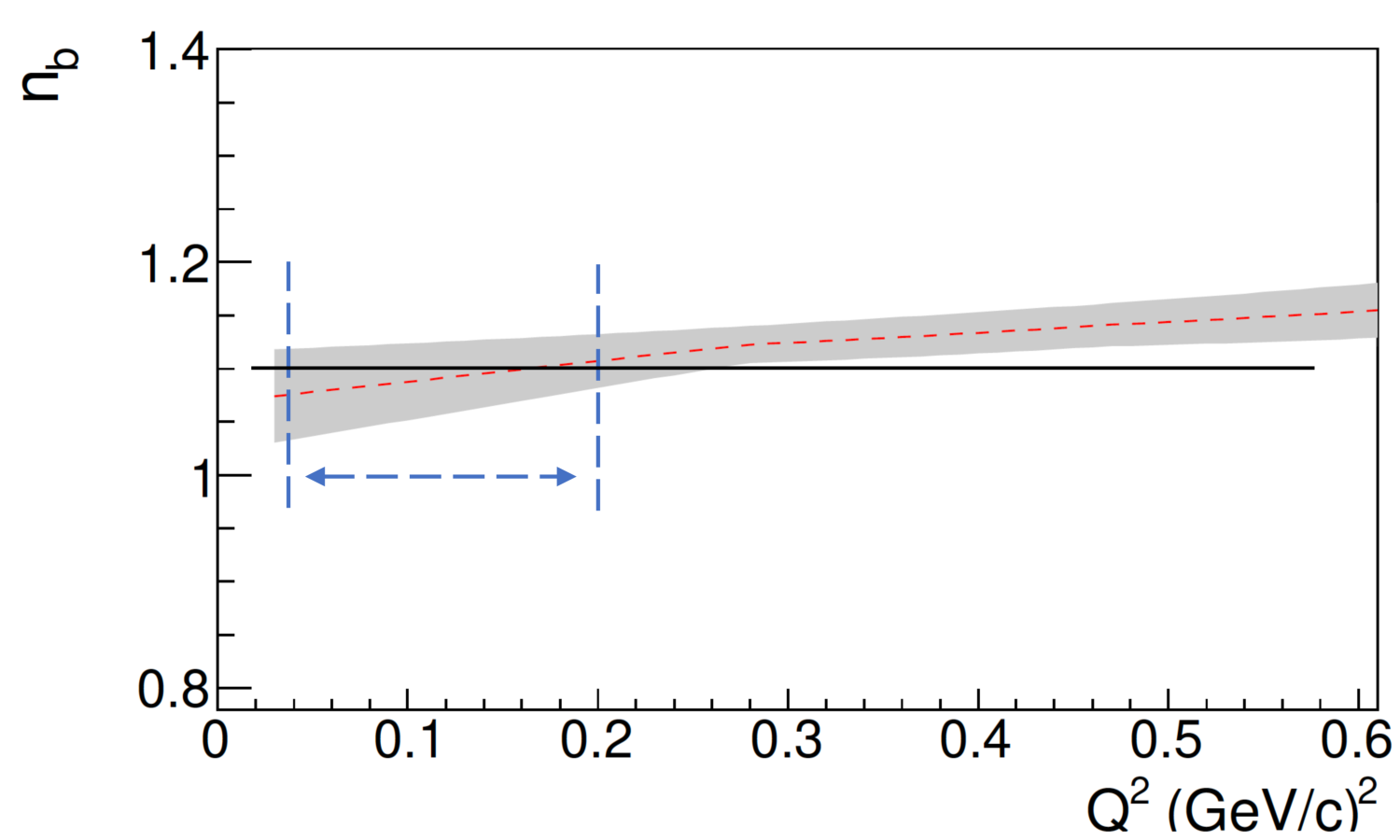}
\caption{The breaking corrections $n_{\rm b}$ (dashed line) and $\delta n_{\rm b}$ uncertainty (shaded band) at the 1$\sigma$ or 68$\%$ confidence level. The solid line indicates the $n_{\rm b}$ as theoretically determined in~\cite{cBuchmann:2004ia}. The horizontal double-arrow marks the $Q^2$-range where the corrections have been employed for the measurement of $G_{\rm E}^{\rm n}$ in this work.}
\label{fig-nb}
\end{figure}

\newpage

\begin{table}[h]
\centering
\begin{tabular}{|llll|}
\hline
$Q^2$  $({\rm GeV}/c)^2$ & $n_{\rm b}$ & $n_{\rm b(min)}$ & $n_{\rm b(max)}$ \\
\hline
\hline
0.040               & 1.076   & 1.034    & 1.119    \\
0.050               & 1.078   & 1.037    & 1.120    \\
0.060               & 1.080   & 1.040    & 1.120    \\
0.070               & 1.082   & 1.043    & 1.121    \\
0.080               & 1.084   & 1.046    & 1.122    \\
0.090               & 1.086   & 1.049    & 1.123    \\
0.100               & 1.088   & 1.052    & 1.124    \\
0.110               & 1.090   & 1.055    & 1.125    \\
0.120               & 1.092   & 1.058    & 1.125    \\
0.130               & 1.093   & 1.061    & 1.126    \\
0.140               & 1.095   & 1.064    & 1.127    \\
0.150               & 1.097   & 1.067    & 1.128    \\
0.160               & 1.099   & 1.070    & 1.129    \\
0.170               & 1.101   & 1.073    & 1.130    \\
0.180               & 1.103   & 1.076    & 1.131    \\
0.190               & 1.105   & 1.079    & 1.132    \\
0.200  	            & 1.107	  & 1.082	 & 1.132    \\
\hline
\end{tabular}
\caption{\label{tabnbc} The $n_{\rm b}$ corrections as experimentally determined in the region of the latest $G_{\rm E}^{\rm n}$ measurements.}
\end{table}

\begin{figure}[t]
\centering
\includegraphics[width=11cm]{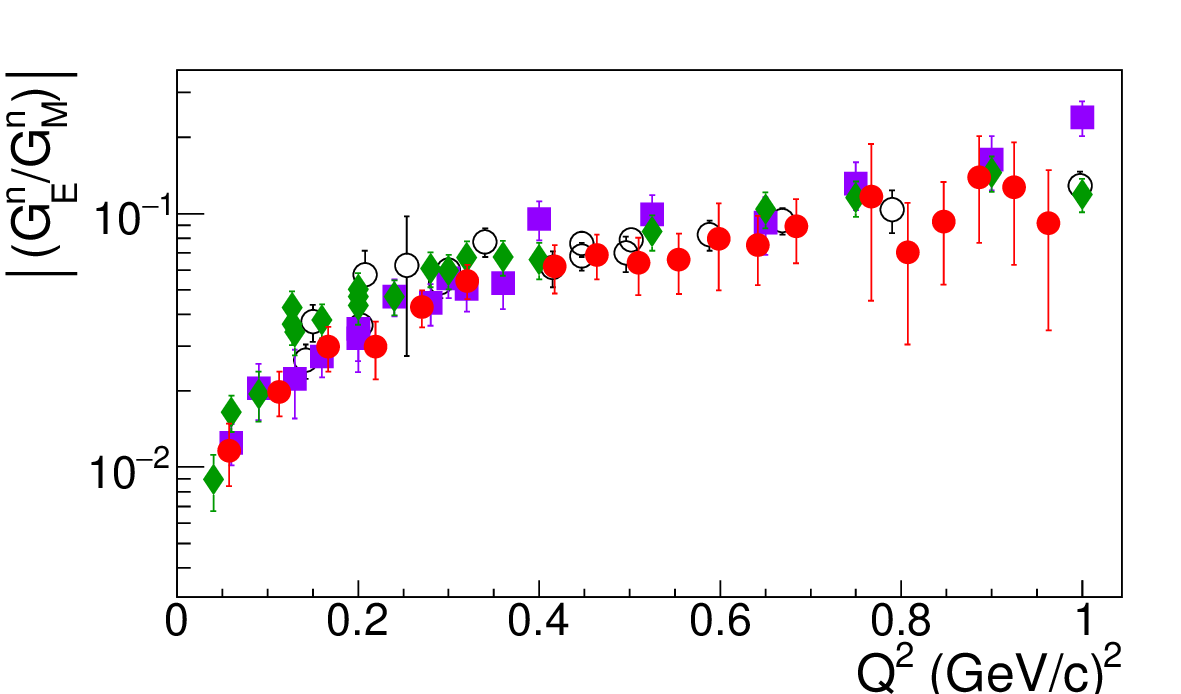}
\caption{{\bf The elastic neutron form factor ratio from the large-$N_{\rm c}$ analysis.} \\The $G_{\rm E}^{\rm n}/G_{\rm M}^{\rm n}$ results from the large-$N_{\rm c}$ analysis with the Coulomb quadrupole data (filled diamonds) and with the Electric quadrupole data (filled boxes) from measurements~\cite{cBlomberg:2015zma,cStave:2006ea,cSparveris:2013ena,cSparveris:2006uk,cSparveris:2004jn,cElsner:2005cz,cAznauryan:2009mx,cKelly:2005jy,cBlomberg:2019caf}. The neutron world data (open-circles) and the LQCD results (filled circles) are also shown. The error bars correspond to the total uncertainty, at the 1$\sigma$ or 68$\%$ confidence level.}
\label{fig-1b}
\end{figure}

\subsection{Analysis within large-$N_{\rm c}$}

The relation between the $G_{\rm E}^{\rm n}$ and the quadrupole transition form factors has also been established through large-$N_{\rm c}$ relations~\cite{cVanderhaeghen:2007}. The relations take the form
\begin{linenomath}\begin{equation}\label{marcemr}
\frac{E2}{M1} (Q^2) =\left (\frac{M_{\rm N}}{M_{\Delta}}\right)^{3/2} \frac{M_{\Delta}^2-M_{\rm N}^2}{2Q^2}\frac{G_{\rm E}^{\rm n}(Q^2)}{F_2^{\rm p}(Q^2)-F_2^{\rm n}(Q^2)}
 \end{equation}\end{linenomath}
 \begin{linenomath}\begin{equation}\label{marccmr}
\frac{C2}{M1} (Q^2) =\left (\frac{M_{\rm N}}{M_{\Delta}}\right)^{3/2} \frac{Q_+Q_-}{2Q^2}\frac{G_{\rm E}^{\rm n}(Q^2)}{F_2^{\rm p}(Q^2)-F_2^{\rm n}(Q^2)}
 \end{equation}\end{linenomath}
where $F_2^{\rm p(n)}$ are nucleon's Pauli form factors, $M_\Delta$ is the mass of the $\Delta$, and $Q_{\pm}=((M_{\Delta} \pm M_{\rm N})^2 + Q^2)^\frac{1}{2}$. 
Here the experimental data base is extended to include the Electric quadrupole (E2) transition, which in turn allows for an improved extraction of the $G_{\rm E}^{\rm n}$. For the well known $G_{\rm E}^{\rm p}$, $G_{\rm M}^{\rm p}$ and $G_{\rm M}^{\rm n}$, that enter in the expressions through the Pauli form factors, we have used recent parametrizations (see next sections for details).

The above relations come with a $15\%$ theoretical uncertainty~\cite{cVanderhaeghen:2007} that has been accounted for in the $G_{\rm E}^{\rm n}$ extraction. That level of uncertainty is further supported by a third relation, also derived in the same work, that connects the isovector Pauli FF to the $N \rightarrow \Delta$ magnetic dipole FF~\cite{cVanderhaeghen:2007}; that relation has been shown to work well at the 13\% level. The level of theoretical uncertainty can be further cross checked using experimental data, when one extracts the $G_{\rm E}^{\rm n}$ independently through the Coulomb and the Electric quadrupole transitions. The results from the two different analyses validate the 15\% level of agreement, as can be seen in Fig.~\ref{fig-1b}. We have thus adopted this number as a theoretical uncertainty in the  data analysis.

From each experiment we derive one $G_{\rm E}^{\rm n}$ value, from both the measurements of the Coulomb and of the Electric quadrupole form factor by taking the weighted average of the two independently derived $G_{\rm E}^{\rm n}$ values. The results are given in Table~\ref{newtable2}. The results are in great agreement with that of the SU(6) analysis. The weighted average of the large-$N_{\rm c}$ and of the SU(6) results leads to the final $G_{\rm E}^{\rm n}$ value. The variance of the two values is assigned as a $G_{\rm E}^{\rm n}$ theoretical uncertainty, and is accounted for accordingly in the $r_{\rm n}$ extraction. The final results for $G_{\rm E}^{\rm n}$ are given in Table~\ref{newtable3}.


\begin{table}[]
\centering
\begin{tabular}{|c|c|c|c|c|}
\hline
$Q^{2}({\rm GeV}/c)^2$ & $G_{\rm E}^{\rm n}$ & $\delta G_{\rm E(exp)}^{\rm n}$ & $\delta G_{\rm E(mod)}^{\rm n}$ & $\delta G_{\rm E(n_{\rm b})}^{n}$\\ \hline
0.040&	0.0128&	0.0011&	0.0034&	0.0005 \\
0.060&	0.0222&	0.0015&	0.0014&	0.0008 \\
0.090&	0.0242&	0.0014&	0.0055&	0.0008 \\ 
0.127&	0.0420&	0.0049&	0.0024&	0.0013 \\
0.130&	0.0388&	0.0015&	0.0065&	0.0012 \\
0.200&	0.0460&	0.0025&	0.0027&	0.0010 \\
0.200&	0.0397&	0.0090&\quad -&	0.0009 \\ \hline
\end{tabular}
\caption{\label{newtable} {\bf The neutron electric form factor derived within the SU(6) analysis.} The $G_{\rm E}^{\rm n}$ results with their experimental and model uncertainties, as described in the text (with the uncertainty $\delta n_{\rm b}$ constrained by the experimental form factor world data).  The last data point, at $Q^2=0.20~({\rm GeV}/c)^2$, involves the measurement from $p(e,e^{'}p)\gamma$~\cite{cBlomberg:2019caf} and it thus does not have a model uncertainty associated with the pion electroproduction background amplitudes.}
\end{table}

\begin{table}[]
\centering
\begin{tabular}{|c|c|c|c|}
\hline
$Q^{2}({\rm GeV}/c)^2$ & $G_{\rm E}^{\rm n}$ & $\delta {G_{\rm E}^{\rm n}}_{\rm (exp)}$ & $\delta {G_{\rm E}^{\rm n}}_{\rm (th)_{1}}$ \\
\hline
0.040&		0.0151&		0.0010&	0.0023 \\
0.060&		0.0246&		0.0011&	0.0037 \\
0.090&		0.0285&		0.0012&	0.0043 \\
0.127&		0.0489&		0.0040&	0.0073 \\
0.130&		0.0447&		0.0013&	0.0067 \\
0.200&		0.0525&		0.0021&	0.0079 \\
0.200&		0.0459&		0.0074&	0.0069 \\
\hline
\end{tabular}
\caption{\label{newtable2} {\bf The neutron electric form factor derived from the large-$N_{\rm c}$ analysis.} The neutron electric form factor derived from the large-$N_{\rm c}$ analysis of the Electric and the Coulomb quadrupole amplitude measurements.}
\end{table}

\begin{table}[]
\centering
\begin{tabular}{|c|c|c|c|}
\hline
$Q^{2}({\rm GeV}/c)^2$ & $G_{\rm E}^{\rm n}$ & $\delta G_{\rm E}^{\rm n}$ & $\delta {G_{\rm E}^{\rm n}}_{\rm (th)_2}$ \\
\hline
0.040&		0.0143&		0.0020&	0.0012 \\
0.060&		0.0228&		0.0019&	0.0012 \\
0.090&		0.0269&		0.0035&	0.0021 \\
0.127&		0.0442&		0.0047&	0.0034 \\
0.130&		0.0417&		0.0048&	0.0030 \\
0.200&		0.0471&		0.0035&	0.0033 \\
0.200&		0.0425&		0.0067&	0.0031 \\

\hline
\end{tabular}
\caption{\label{newtable3} {\bf Final results for the neutron electric form factor.} The $G_{\rm E}^{\rm n}$ final results as derived from the SU(6) and the large-$N_{\rm c}$ analysis. The theoretical uncertainty is derived from the variance of the results from the two analyses.}
\end{table}


\section{Neutron charge radius extraction}

\begin{figure}[ht]
\centering
\includegraphics[width=\textwidth]{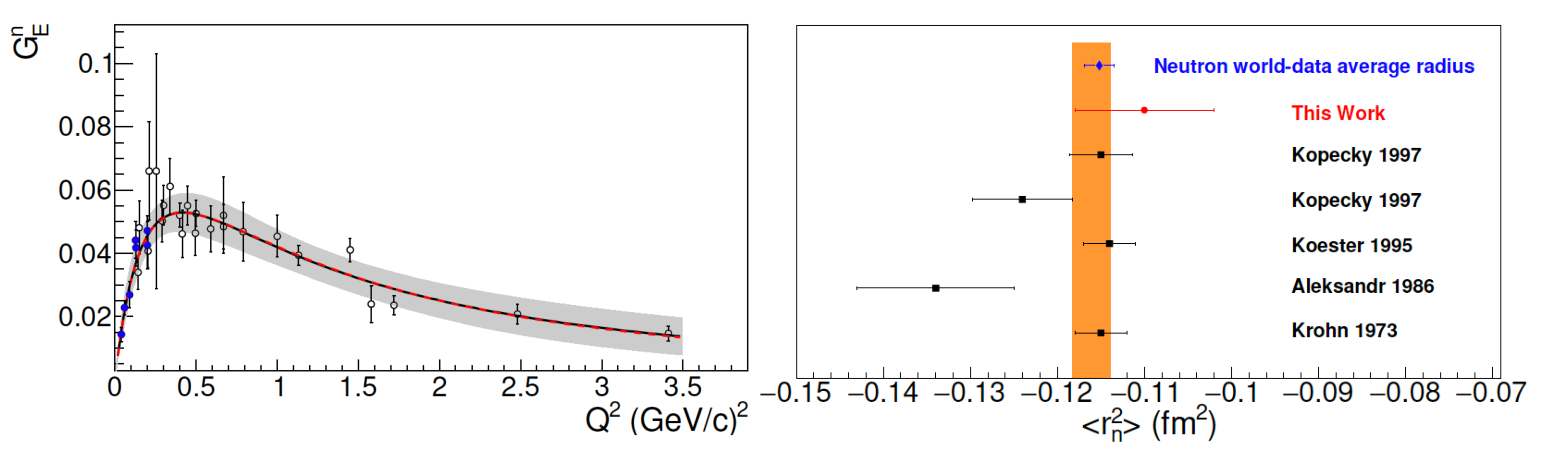}
\caption{Left panel: The $G_{\rm E}^{\rm n}$ results from this work (filled-circles) and the $G_{\rm E}^{\rm n}$ world data (open-circles)~\cite{cMadey:2003av,cSchlimme:2013eoz,cRiordan:2010id,cGlazier:2004ny,cPlaster:2005cx,cZhu:2001md,cWarren:2003ma,cRohe:1999sh,cBermuth:2003qh,cGeis:2008aa,cEden.50.R1749,costrick.83.276,cgolak.63.034006,cHerberg:1999ud}. The error bars correspond to the total uncertainty, at the 1$\sigma$ or 68$\%$ confidence level. The solid (black) curve shows the fit to the data from the parametrization of Eq.~\ref{modgalster} with its uncertainty (shaded band). The dashed (red) line shows the fit from parametrization of Eq.~\ref{2dipolfit}. The two fits practically overlap. Right panel: The $\langle r_{\rm n}^2 \rangle$ measurement from this work and from the references~\cite{cKopecky:1997rw,cKoester:1995nx,cAleksandrov:1986mw,cKrohn:1973re} that are currently included in the PDG $\langle r_{\rm n}^2 \rangle$ analysis. The orange-band marks the PDG averaged $\langle r_{\rm n}^2 \rangle$ value. The recent weighted average of the world data is also shown, when the $\langle r_{\rm n}^2 \rangle$ measurement reported here is included in the calculation.}
\label{fig-radius}
\end{figure}

The neutron mean square charge radius is related to the slope of the neutron electric form factor as $Q^2 \rightarrow 0$ through 
\begin{linenomath}\begin{equation}\label{slope} \langle r_{\rm n}^2 \rangle = \left.-6 \frac{dG_{\rm E}^{\rm n}(Q^2)}{dQ^2} \right\rvert_{Q^2 \rightarrow 0}. \end{equation}\end{linenomath}
The $G_{\rm E}^{\rm n}(Q^2)$ has to be parametrized and fitted to the experimental data, and from the slope at $Q^2=0$ the $\langle r_{\rm n}^2 \rangle$ is determined. The data derived in this work offer the level of precision, and the $Q^2$-range, that allow to explore for suitable $G_{\rm E}^{\rm n}(Q^2)$ functional forms with an additional free parameter compared to what was possible in the past (i.e. with three free parameters). In turn, this allows to extract the $\langle r_{\rm n}^2 \rangle$ from the $G_{\rm E}^{\rm n}$ measurements, which has not done before. We have explored various functional forms in order to identify those appropriate, in the following way:\\ 
i) The functions have been fitted to the $G_{\rm E}^{\rm n}$ results reported in this work (data-set in Table~\ref{newtable3}) and to the world data~\cite{cMadey:2003av,cSchlimme:2013eoz,cRiordan:2010id,cGlazier:2004ny,cPlaster:2005cx,cZhu:2001md,cWarren:2003ma,cRohe:1999sh,cBermuth:2003qh,cGeis:2008aa,cEden.50.R1749,costrick.83.276,cgolak.63.034006,cHerberg:1999ud}. \\
ii) For the functions that offer a good fit to the data, further tests are performed with pseudo-data; these tests aim to place the parametrizations under stress in terms of their limitations to fits that engage data of higher precision and of extended kinematical coverage. \\
iii) The $\langle r_{\rm n}^2 \rangle$ extraction is repeated with multiple $G_{\rm M}^{\rm n}$ parametrizations so that the uncertainty introduced by the $G_{\rm M}^{\rm n}$ parametrization is quantified; for that part we have found $\pm 0.0009~({\rm fm}^2)$, an order of magnitude smaller compared to the total $\langle r_{\rm n}^2 \rangle$ uncertainty.\\
Our studies have shown that the most robust function for the radius extraction takes the form
\begin{linenomath}\begin{equation}\label{modgalster} G_{\rm E}^{\rm n}(Q^2)=(1 + Q^2/A)^{-2} \frac{B \tau}{1 + C \tau} , \end{equation}\end{linenomath}
where $\tau=Q^2/4m^2$, and $A, B, C$ are free parameters. It involves a similar form to the Galster~\cite{cGalster:1971kv}. The Galster is a long standing phenomenological parametrization that could adequately describe the early $G_{\rm E}^{\rm n}$ data, but as recent experiments revealed~\cite{gentile:2011}~\footnote{The reference Phys. Rev. C83, 055203 (2011) was inadvertently not included in the version of the paper that was submitted at Nature Communications} it does not have sufficient freedom to accommodate reasonable values of the radius, without constraining or compromising the fit (note: the reference Phys. Rev. C83, 055203 (2011) was inadvertently not included in the version of the paper that was submitted at Nature Communications). Here, instead of using the standard dipole form factor with $\Lambda^2=0.71({\rm GeV}/c)^2$ an additional free parameter $A$ is introduced (see Eq.~\ref{modgalster}). Our fit to the data gives $\langle r_{\rm n}^2 \rangle = -0.110 \pm0.008~({\rm fm}^2)$ with a reduced $\chi^2$ of 0.74. The fitted parameters are given in Table~\ref{radparam}.\\ 
A second parametrization, giving a good fit to the data, involves the sum of two dipoles
\begin{equation}\label{2dipolfit} G_{\rm E}^{n}(Q^2)= \frac{A}{(1+\frac{Q^{2}}{B})^2} -\frac{A}{(1+\frac{Q^{2}}{C})^2}. \end{equation}
This form has been explored in the past~\cite{cGeis:2008aa,gentile:2011} with three free parameters and with the $\langle r_{\rm n}^2 \rangle$ already constrained by the measurement of the neutron-electron scattering length. This fit (see Table~\ref{radparam}) exhibits an excellent agreement to the one of Eq.~\ref{modgalster}. The two curves are nearly indistinguishable by eye as seen in Fig.~\ref{fig-radius}a. Considering the weighted average of the two methods we derive a nearly identical result of $\langle r_{\rm n}^2 \rangle = -0.110 \pm0.007~({\rm fm}^2)$. Nevertheless, our pseudo-data studies showed that the two-dipole fit suffers from limitations in the determination of the $G_{\rm E}^{\rm n}$-slope. We thus adopt only the paramterization of Eq.~\ref{modgalster} and quote it's fitted result $\langle r_{\rm n}^2 \rangle = -0.110 \pm0.008~({\rm fm}^2)$ as our final result for the neutron charge radius.

\begin{table}[h]
\centering
\begin{tabular}{|l|l|l|l|}
\hline
Parametrization & A & B   & C \\ \hline
Param.~\ref{modgalster} & 0.505 $\pm$ 0.079 & 1.655 $\pm$ 0.126 &  0.909 $\pm$ 0.583 \\
\hline
Param.~\ref{2dipolfit}    & 0.130 $\pm$ 0.039  & 1.790$ \pm$ 0.409   &  0.419 $\pm$ 0.0980 \\
\hline
\end{tabular}
\caption{\label{radparam} Fitted parameters for the two $G^{\rm n}_{\rm E}$ parametrizations.}
\end{table}

We have also explored the radius extraction by adopting the scenario where the uncertainty of the symmetry breaking corrections in the SU(6) analysis is treated very conservatively i.e. $n_{\rm b}=1.1 \pm 0.1$. In this case the final result becomes $\langle r_{\rm n}^2 \rangle = -0.109 \pm0.009~({\rm fm}^2)$ with a reduced $\chi^2$ of 0.74. This scenario tends to overestimate the uncertainty of the symmetry breaking terms, but as this is not a dominant factor in the radius extraction the $r_{\rm n}$-uncertainty is not affected significantly.

The particle data group (PDG) average value, $\langle r_{\rm n}^2 \rangle = -0.1161 \pm0.0022~({\rm fm}^2)$, is the weighted average of five values~\cite{cKopecky:1997rw,cKoester:1995nx,cAleksandrov:1986mw,cKrohn:1973re} based on the measurement of the neutron-electron scattering length. One of these measurements~\cite{cAleksandrov:1986mw} disagrees with ~\cite{cKopecky:1997rw,cKoester:1995nx}, but as this disagreement has not been resolved the value from measurement~\cite{cAleksandrov:1986mw} is considered in the PDG average, for the time being. Our result, $\langle r_{\rm n}^2 \rangle = -0.110 \pm0.008~({\rm fm}^2)$, comes to update the world average value for the neutron charge radius. The recent measurement, like the ones from~\cite{cKopecky:1997rw,cKoester:1995nx}, also disagrees with that of ~\cite{cAleksandrov:1986mw}, and considering that this disagreement comes from a different method for the radius extraction one may want to exclude the value of~\cite{cAleksandrov:1986mw} from the world data average. The latest weighted average value, including our measurement and excluding the one of~\cite{cAleksandrov:1986mw}, becomes $\langle r_{\rm n}^2 \rangle = -0.1152 \pm0.0017~({\rm fm}^2)$ and is shown in Fig.~\ref{fig-radius}b. In the current PDG world-data average, the uncertainty that is calculated following the standard weighted least-squares procedure is further enhanced by a factor of $S=[\chi^2/(N-1)]^{1/2}$ due to the discrepancies that are introduced from the inclusion of the Alexandrov'86 measurement in the data-set. In the most recent world-data average, with the inclusion of the new measurement and the subsequent exclusion of the Alexandrov'86, these discrepancies are raised and the term $S$ does not contribute any further to the world-data average uncertainty. Effectively, the biggest benefit to the uncertainty results from the resolution of the discrepancies in the $\langle r_{\rm n}^2 \rangle$ measurements that the current work allows for. For the completeness of this discussion, we can also comment that if one considers in the world-data average both the new $\langle r_{\rm n}^2 \rangle$ measurement as well as that of Alexandrov'86, then the uncertainty improves by 10\%, from $\pm0.0022~({\rm fm}^2)$ to $\pm0.0020~({\rm fm}^2)$. Nevertheless, the results of this work suggest that the Alexandrov'86 has to be eliminated from the world-data average in which case the uncertainty is further reduced to $\pm0.0017~({\rm fm}^2)$.

\subsection{Extraction from fits within a limited low-$Q^2$ range}

The extraction of the neutron charge radius has been performed by fitting over an extended $Q^2$ range, following the experience from the extraction of the proton charge radius, where the world data base is far more extensive compared to the one of the neutron. In that case, the charge radius extraction has been traditionally based on fits of functional forms over an extended kinematic $Q^2$ range (e.g. such as in the most recent $r_p$ extraction from the Mainz $G_{\rm E}^{\rm p}$ measurements, etc). Fitting over a limited low~$Q^2$-range in order to extract the proton radius was first attempted very recently, at the PRad experiment. In that case though the $G_{\rm E}^{\rm p}$ was accessed at much lower momentum transfers compared to the neutron measurements, namely in the range $Q^2=0.0002~({\rm GeV}/c)^2$ to $Q^2=0.06~({\rm GeV}/c)^2$.

\begin{figure}[t]
\centering
\includegraphics[width=\textwidth]{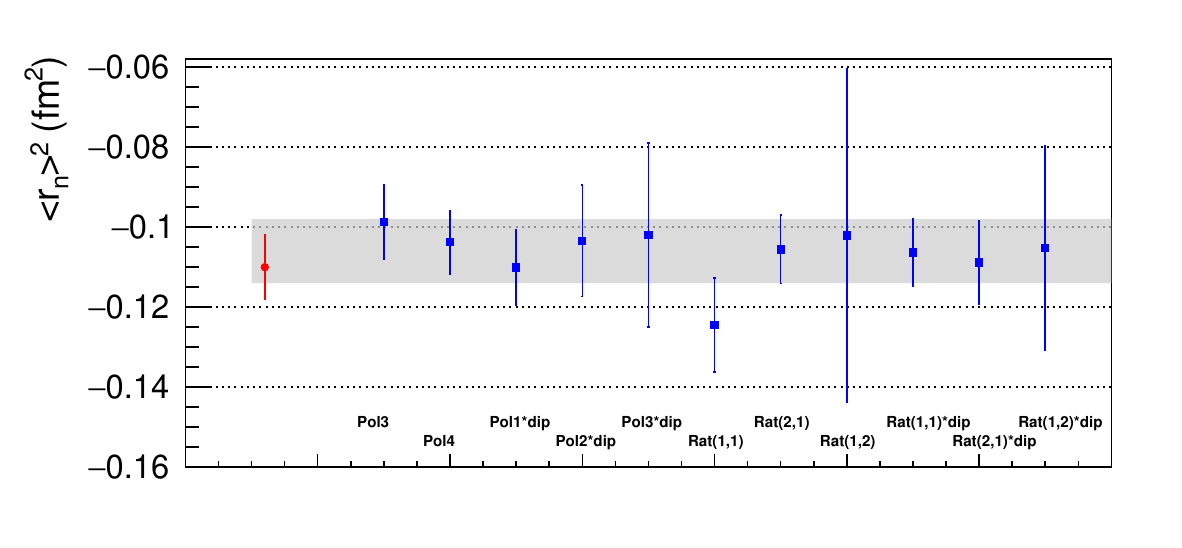}
\caption{The $\langle r_{\rm n}^2 \rangle$ as extracted with the different model functions (blue boxes) from fits within the low $Q^2$ range. The error bars correspond to the total uncertainty, at the 1$\sigma$ or 68$\%$ confidence level. The gray band marks the final result, from the combination of all the fits of both groups of functions (see text for details as well as Tab.~\ref{tablowqfits}). The red point marks the final result from the fit over the extended $Q^2$ range, namely $\langle r_{\rm n}^2 \rangle = -0.110 \pm0.008~({\rm fm}^2)$.}
\label{figlowqfit}
\end{figure}

\begin{figure}[t]
\centering
\includegraphics[width=10.0cm]{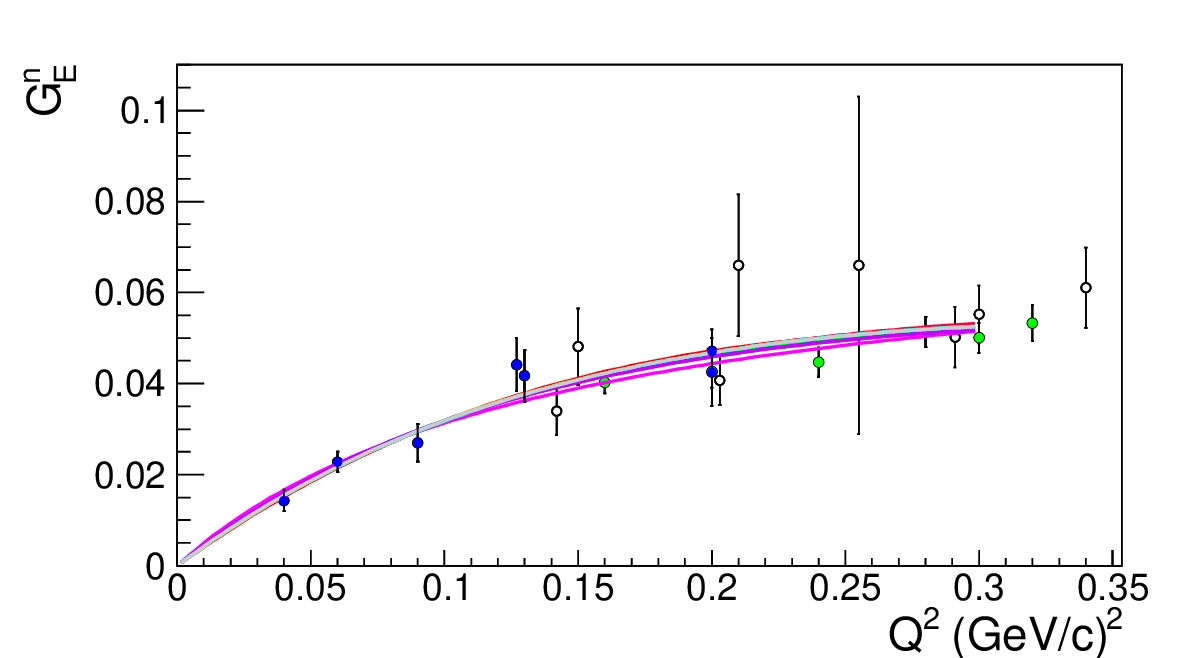}
\caption{The low-$Q^2$ $G_{\rm E}^{\rm n}$ fits with all the functions within the polynomial and the rational groups. The data from the analysis of the MAMI and Hall-A measurements are marked as blue points and the ones from the CLAS as green points. The error bars correspond to the total uncertainty, at the 1$\sigma$ or 68$\%$ confidence level.}
\label{figlowqgenfit}
\end{figure}

One consideration here is that the proton and the neutron electric form factor parametrizations are inherently different (i.e. the $G_{\rm E}^{\rm n}$ changes slope and follows a monotonic fall-off at low $Q^2$). With that in mind we here explore the potential of $G_{\rm E}^{\rm n}$ fits that are limited only in the low $Q^2$ region. The global fit for $G_{\rm E}^{\rm n}$ shows that a maximum is reached shortly after $Q^2=0.4~({\rm GeV}/c)^2$. We thus limit our studies within the range [0 - $0.4~({\rm GeV}/c)^2$] that $G_{\rm E}^{\rm n}$ is expected to be monotonic. In doing so one has to consider to following factors: what is the choice of the functional forms and the number of parameters that will be fitted, what is the optimal fitting range, and how does the extracted radius depend on the ansatz for the fit model. Here we follow a procedure that has been frequently adopted for the extraction of the proton charge radius. We have studied a variety of functions so that we can identify those that are appropriate for the fit to the data. The functional forms that allow the extraction of the charge radius with a meaningful precision are divided in two groups. The first group is based on polynomials and involves polynomials with varying orders and combinations of polynomials with a dipole. The stability of the extracted radius is observed within the group by using polynomials of different degree, while the variance of the fitted values indicates the model uncertainty of the group. The second group is based on rational functions of the form $$Rat(i,j)(Q^2)=\frac{\sum\limits_{i=1}^{n} \alpha_i Q^{2i}}{1 + \sum\limits_{j=1}^{m} \beta_j Q^{2j}}$$ and combinations of Rat(n,j) with dipole. These groups of functions have been utilized in the past for the proton radius extraction. Other functional forms that we explored (e.g. exponential-based functions $f(Q^2)=\sum\limits_{i=0}^{n} \alpha_i Q^{2i}(1 - e^{-\beta Q^{2}})$, etc) give consistent results for the charge radius but with large uncertainties and as such we do not consider them here. When it comes to the number of fitted parameters, $n_p$, one has to make sure that $n_p$ is not too small since in such a case the data will not be properly reproduced, while at the same time one has to avoid a very high level of flexibility in order to avoid erratic fits. For example, in the polynomial fits we have found that starting with a 3rd order polynomial we can have a good fit, but after the 5th order polynomial the fitted function follows an erratic, non-monotonic behavior through the data points at the higher end of the $Q^2$ fitting range. In the latter case, the fitted value for the $\langle r_{\rm n}^2 \rangle$ is found to be consistent to that of the fits with the lower order polynomials. Nevertheless, for consistency we have decided to exclude such fits from the determination of the charge radius since our requirement is to fit a monotonic function at low momentum transfers. Polynomials$\times$dipole were able to provide a good fit, contrary to polynomial+dipole that would result into non-stable, erratic fits. Functional forms within the group were not considered when their resulting uncertainties were too large to contribute meaningfully to the determination of the charge radius, such as rational functions higher than Rat(2,2). The $\langle r_{\rm n}^2 \rangle$ results from the individual fits within the two groups are shown in Fig.~\ref{figlowqfit}. The comparison of all the $G_{\rm E}^{\rm n}$ fits for the functions within the two groups are shown in Fig.~\ref{figlowqgenfit}.

\begin{table}[h!]
\centering
\begin{tabular}{|l|l|l|}
\hline
      & [0 - $0.3~({\rm GeV}/c)^2$] & [0 - $0.4~({\rm GeV}/c)^2$]  \\ \hline
Polynomial group & $\langle r_{\rm n}^2 \rangle = -0.105 \pm0.006 \pm0.002_{\rm mod}~({\rm fm}^2)$ & $\langle r_{\rm n}^2 \rangle = -0.104 \pm0.005 \pm0.003_{\rm mod}~({\rm fm}^2)$ \\
Rational group & $\langle r_{\rm n}^2 \rangle = -0.108 \pm0.006 \pm0.002_{\rm mod}~({\rm fm}^2)$ & $\langle r_{\rm n}^2 \rangle = -0.108 \pm0.006 \pm0.002_{\rm mod}~({\rm fm}^2)$ \\
\hline
 \multicolumn{3}{|c|}{$\langle r_{\rm n}^2 \rangle = -0.107 \pm0.006 \pm0.002_{\rm mod} \pm0.002_{\rm group}~({\rm fm}^2)$} \\
\hline
\end{tabular}
\caption{\label{tablowqfits} The extracted $\langle r_{\rm n}^2 \rangle$ from the fits within a low-$Q^2$ range.}
\end{table}

For the charge radius, the weighted average is extracted separately for each one of the two groups. A systematic uncertainty is also quantified within each group (i.e. a model uncertainty of the group) from the weighted variance of the results from all the fits within the group. The results from the two groups tend to have a similar total uncertainty (i.e. when the weighted average and the group's model uncertainty are combined). Nevertheless there is a small systematic difference of the two group's central values, as studies over a varying fitting range have shown. For that reason a third uncertainty is determined: here we consider the spread of the two central values as indicative of the uncertainty that is associated with the choice of the group. Therefore the final result is given by the average of these two values, while the half of the difference of the two values is assigned as an additional (group) uncertainty (see Tab.~\ref{tablowqfits}). This uncertainty has to be added linearly, and not quadratically, to the other two uncertainties. Lastly, the sensitivity of the results to the fitting range has been explored. There is an interplay of the fitting and of the model/group uncertainties depending on the fitting range in $Q^2$. Our studies have shown that as we increase the upper bound of the fitting range, $Q^2_{\rm max}$, the fitting uncertainties tend to improve but at the expense of the group's model uncertainty. When all uncertainties are considered, the overall level of uncertainty for the results within the [0 - $0.3~({\rm GeV}/c)^2$] and within the [0 - $0.4~({\rm GeV}/c)^2$] range are comparable, with a very small benefit if the fits are conducted within the [0 - $0.3~({\rm GeV}/c)^2$]. As a general remark, when the overall uncertainty (i.e. from the combination of the fitting and model ones) is equivalent for two different fitting ranges it may be preferable to pursue the results through fits within the range where the model uncertainties tend to minimize, considering that the fitting uncertainties are very strictly defined and accurately determined, while the determination of model uncertainties entails a higher level of uncertainty. Lastly, limiting the fitting range even further, within the [0 - $0.2~({\rm GeV}/c)^2$] range, is not beneficial since the fitting uncertainties increase significantly. The results from the fits within each group are shown in Tab.~\ref{tablowqfits}. With the low-$Q^2$ fits we find that $\langle r_{\rm n}^2 \rangle = -0.107 \pm0.006 \pm0.002_{\rm mod} \pm0.002_{\rm group}~({\rm fm}^2)$. This value is in excellent agreement compared to what we derive when we fit within the extended $Q^2$ range, namely $\langle r_{\rm n}^2 \rangle = -0.110 \pm0.008~({\rm fm}^2)$. Nevertheless we conclude that the fits at low-$Q^2$ are not able to provide a more precise determination of $\langle r_{\rm n}^2 \rangle$ compared to the fit over the extended $Q^2$ range, while a higher level of model uncertainty is introduced to the result by the low-$Q^2$ fits. We here note that the inclusion of the CLAS data in the low-$Q^2$ fits (see Section 3.2 that follows) reveal a further increase to the {\it group} uncertainty, from $\pm0.002_{\rm group}~({\rm fm}^2)$ to $\pm0.004_{\rm group}~({\rm fm}^2)$. Our studies indicate that more data are required lower in $Q^2$ so that these uncertainties can be further suppressed, and that the $G_{\rm E}^{\rm n}$ fits over the extended $Q^2$ range provide the most precise and consistent measurement of $\langle r_{\rm n}^2 \rangle$.

\subsection{Extraction including additional data at higher $Q^2$}

The limiting factor in the extraction of the neutron charge radius has been the lack of $G_{\rm E}^{\rm n}$ data at low momentum transfers. This limitation is now raised by the latest data provided by the MAMI and the JLab/Hall-A measurements in the region $Q^2=0.04~({\rm GeV}/c)^2$ to $Q^2=0.20~({\rm GeV}/c)^2$. The intermediate and the high~$Q^2$ regime on the other hand is already sufficiently covered by the $G_{\rm E}^{\rm n}$ world data and the $\langle r_{\rm n}^2 \rangle$ fits are not limited by the luck of data in this region.

\begin{figure}[h!]
\centering
\includegraphics[width=13.0cm]{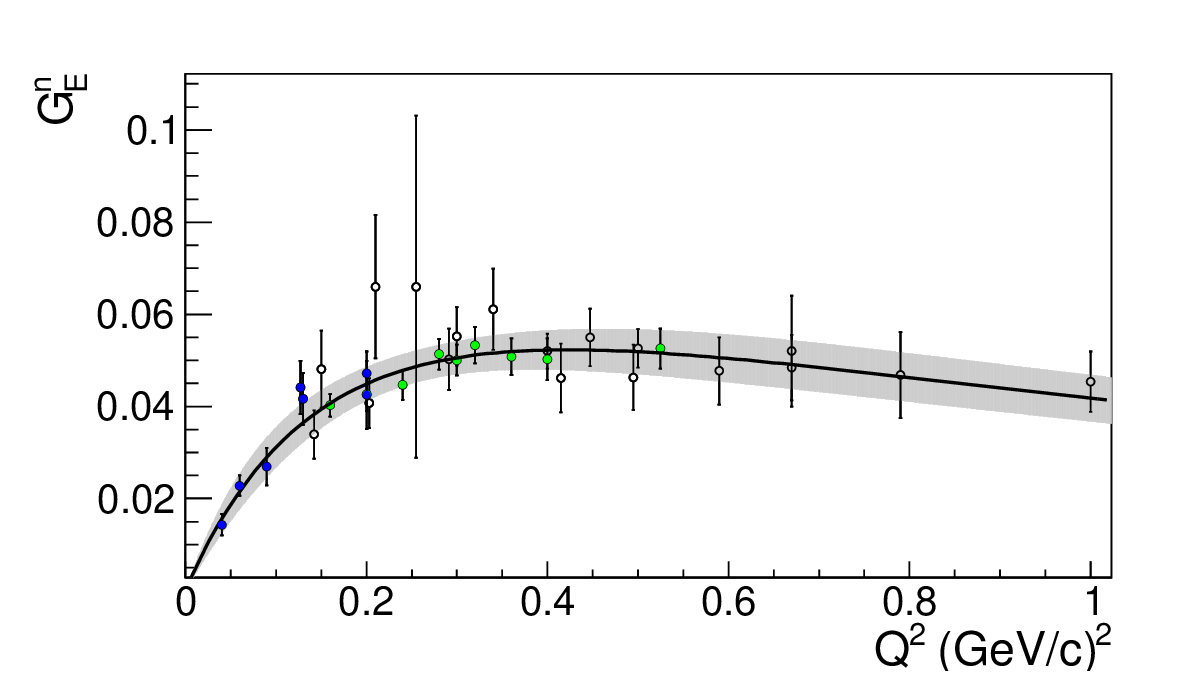}
\caption{The $G_{\rm E}^{\rm n}$ results from the MAMI and the JLab/Hall-A measurements (blue points) and from the CLAS measurements (green points). The error bars correspond to the total uncertainty, at the 1$\sigma$ or 68$\%$ confidence level.}
\label{figgenclas}
\end{figure}

The quadrupole transition form factor measurements extend higher in $Q^2$, in the region that has already been accessed by $G_{\rm E}^{\rm n}$ measurements. One can thus naturally consider to extend the current analysis to higher momentum transfers so as to enrich the $G_{\rm E}^{\rm n}$ data base even further, with overlapping measurements, in the hope to improve the fits for the $\langle r_{\rm n}^2 \rangle$ extraction. In principle one can attempt to do that. Nevertheless, while the relations that relate the $G_{\rm E}^{\rm n}$ to the quadrupole transition form factors hold on very solid ground in the low~$Q^2$ region (i.e. the region of the MAMI and the JLab/Hall-A measurements) they tend to hold less well at high momentum transfers. There is no sharp, formally defined $Q^2$ cut-off-value after which these relations do not hold, and as such they should be treated with caution at increasing momentum transfers e.g. their level of theoretical uncertainty tend to increase around $\approx 1~({\rm GeV}/c)^2$ to a level that one may not wish to risk including them in the current analysis, considering the level of precision that we are after for the charge radius extraction. One thus has to walk along a fine line, by adding measurements at intermediate momentum transfers, e.g. higher than $Q^2=0.20~({\rm GeV}/c)^2$ but not going very high in $Q^2$, so that a clear benefit to the $r_{\rm n}$ extraction is achieved and without compromising the extraction with theoretical uncertainties.

Here we expand the range of our analysis and we extract the $G_{\rm E}^{\rm n}$ from the JLab/CLAS~\cite{cAznauryan:2009mx} lowest momentum-transfer data-set  which involves measurements of the quadrupole transition form factors up to $Q^2=0.52~({\rm GeV}/c)^2$. The measurements are taken within a kinematic region where one can still feel comfortable applying these relations. The extracted $G_{\rm E}^{\rm n}$ from these measurements is shown in Fig.~\ref{figgenclas}, and the results are in excellent agreement with the $G_{\rm E}^{\rm n}$ world data in the same region. When we include this additional set of data in the $\langle r_{\rm n}^2 \rangle$  extraction and we fit over the complete $Q^2$-range of the $G_{\rm E}^{\rm n}$ measurements we find a  $\langle r_{\rm n}^2 \rangle = -0.107 \pm0.007~({\rm fm}^2)$, compared to $\langle r_{\rm n}^2 \rangle = -0.110 \pm0.008~({\rm fm}^2)$ when the CLAS data are not included. Here we observe a minor improvement to the $\langle r_{\rm n}^2 \rangle$ uncertainty with the inclusion of the CLAS data, but we suggest that this result is treated with some caution. We have some concerns, that part of the reported CLAS uncertainties may be underestimated which could in turn influence / bias the level of the $\langle r_{\rm n}^2 \rangle$ uncertainty. We  discuss these considerations in more detail in the section that follows (Section 3.2.1).

We do not consider extending the data-set with measurements at higher $Q^2$ e.g. by including the CLAS data that are higher than $Q^2 \approx 0.7~({\rm GeV}/c)^2$, for the following reasons. Firstly, as we noted above, in such a case we would be entering a region that is associated with a high level of theoretical uncertainties. Secondly, when we investigate the effect of including the additional data up to $Q^2=1~({\rm GeV}/c)^2$ we did not observe any further benefit to the fits, namely we derive the same level of $\langle r_{\rm n}^2 \rangle$ uncertainty. Thus in such a case one would risk to introduce further uncertainties of theoretical nature to the $\langle r_{\rm n}^2 \rangle$ extraction without any benefit to the fitting uncertainty. Lastly, the CLAS results at $Q^2=1~({\rm GeV}/c)^2$ disagree with the  results from the Hall-A experiment~\cite{cKelly:2005jy} at the same $Q^2$ that is based on recoil polarization measurements. The Hall-A experiment offers a nearly model-independent analysis and also utilizes a setup with superior resolution compared to the one of CLAS. If we were to consider these (disagreeing) data we could risk further bias to the $\langle r_{\rm n}^2 \rangle$ extraction from the tensions among the competing transition form factors measurements in that region.

\begin{figure}[h!]
\centering
\includegraphics[width=14.0cm]{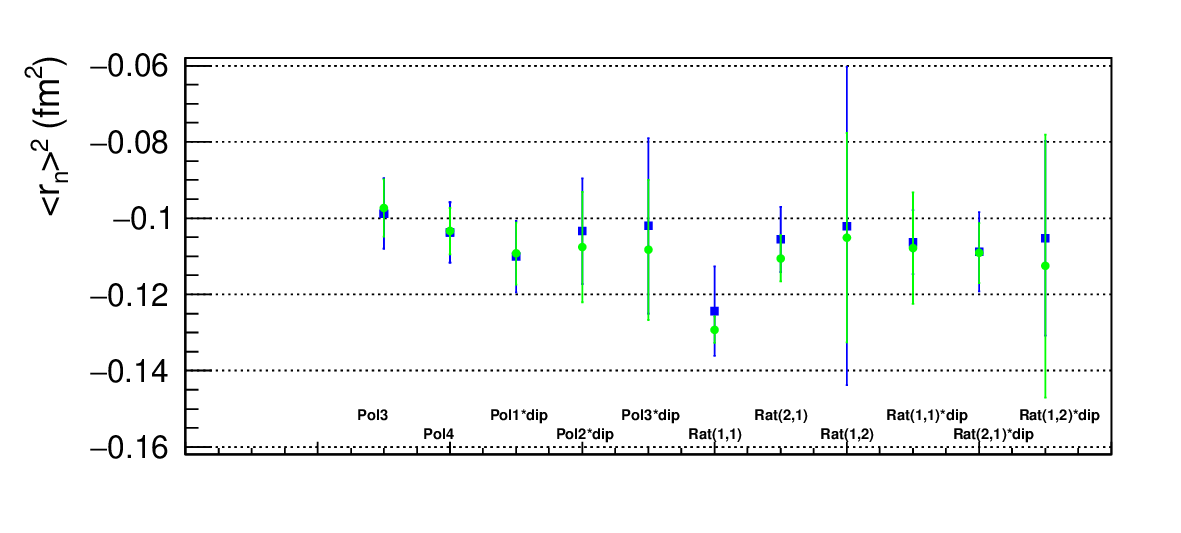}
\caption{The $\langle r_{\rm n}^2 \rangle$ as extracted with the different model functions from fits within the low $Q^2$ range. The results are shown with the inclusion of the CLAS data (green circles) and without (blue boxes). The error bars correspond to the total uncertainty, at the 1$\sigma$ or 68$\%$ confidence level.}
\label{figfitcomparison}
\end{figure}

Lastly, we explore the influence of the CLAS data to the $\langle r_{\rm n}^2 \rangle$ extraction from the fits within a limited low-$Q^2$ range, as discussed in Section 3.1. The results for the fitted functions when the CLAS data are included are shown in Fig.~\ref{figfitcomparison} and they are compared to the results of the fits when the CLAS data are not included in the analysis. The final results are summarized in Tab.~\ref{tablowqfitsclas}. Looking at the results we are able to make a number observations and to draw a set of conclusions with regard to the strength of the low-$Q^2$ fits. First, we note that the results of the polynomial group maintains remarkable stability compared to the analysis when the CLAS data are not included. The central value of the rational group, on the other hand, observes a shift compared to the analysis when the CLAS data are not included. This effectively enhances the group uncertainty from $\pm0.002_{\rm group}~({\rm fm}^2)$ to $\pm0.004_{\rm group}~({\rm fm}^2)$ when the CLAS data are included. Another observation is that the model uncertainty within each group tends to also increase when we fit in the [0 - $0.4~({\rm GeV}/c)^2$] range, but not at the [0 - $0.3~({\rm GeV}/c)^2$] range. One can conclude that for low-$Q^2$ fits within that region we are rather sensitive to model uncertainties that depend on the choice of the fitted parametrization. These uncertainties tend to be comparable to the statistical uncertainty of the fits. These uncertainties can be further suppressed when more data are acquired at lower momentum transfers and the fitting range can be limited even lower in $Q^2$.

\begin{table}[h!]
\centering
\begin{tabular}{|l|l|l|}
\hline
      & [0 - $0.3~({\rm GeV}/c)^2$] & [0 - $0.4~({\rm GeV}/c)^2$]  \\ \hline
Polynomial group & $\langle r_{\rm n}^2 \rangle = -0.107 \pm0.006 \pm0.001_{\rm mod}~({\rm fm}^2)$ & $\langle r_{\rm n}^2 \rangle = -0.104 \pm0.004 \pm0.004_{\rm mod}~({\rm fm}^2)$ \\
Rational group & $\langle r_{\rm n}^2 \rangle = -0.115 \pm0.006 \pm0.002_{\rm mod}~({\rm fm}^2)$ & $\langle r_{\rm n}^2 \rangle = -0.115 \pm0.005 \pm0.007_{\rm mod}~({\rm fm}^2)$ \\
\hline
 \multicolumn{3}{|c|}{$\langle r_{\rm n}^2 \rangle = -0.111 \pm0.006 \pm0.002_{\rm mod} \pm0.004_{\rm group}~({\rm fm}^2)$} \\
\hline
\end{tabular}
\caption{\label{tablowqfitsclas} The extracted $\langle r_{\rm n}^2 \rangle$ from the fits within a low-$Q^2$ range when the CLAS measurements are included in the data-set. The final result, from the average of the two groups, is given for the fits in the [0 - $0.3~({\rm GeV}/c)^2$] range, where the model uncertainties are minimized.}
\end{table}

\subsubsection{Considerations on the CLAS data uncertainties} 

The CLAS results agree nicely with the MAMI and JLab/Hall-A measurements within their region of overlap at $Q^2 \approx 0.2~({\rm GeV}/c)^2$. Nevertheless, we have some concerns that part of the CLAS uncertainties may be underestimated. If that is indeed the case, one could risk to bias the $\langle r_{\rm n}^2 \rangle$ extraction when these data are included in the analysis. More specifically, the CLAS detector has a large acceptance coverage (4$\pi$), which offers some advantage as to the extent of the measured phase space. Nevertheless, this advantage comes at the expense of the detector's resolution, when compared to the high resolution spectrometers that have been utilized for the MAMI and the JLab/Hall-A measurements. In the latter case (i.e. at MAMI and in Hall-A) the extended phase space is still covered by sequential measurements of the high resolution spectrometers so that one can ultimately achieve an extended phase space coverage that is comparable to the CLAS one. Thus one would naturally expect that ultimately the MAMI and the JLab/Hall-A results will offer smaller overall experimental uncertainties in the transition form factor extraction since e.g. the expectation is that the limitations in the resolution of CLAS will introduce a higher level of systematic uncertainties in the CLAS results. But this is not what one observes in the reported results. The CLAS measurements are reported with superior experimental uncertainties (compared to the MAMI and the Hall-A data). This points out to a potential underestimation for part of the experimental uncertainties within the global fit analysis of the extended CLAS 4$\pi$ acceptance. We here note that the extraction of the quadrupole transition amplitudes involves a very delicate task since it requires to isolate and extract a very small amplitude from multiple interfering background processes of similar magnitude. A second concern involves the level of the model uncertainties of the CLAS data. One would naturally expect that these uncertainties are not very much dependent on the experimental setup, and that they will be approximately at the same level for all data sets (CLAS, Hall-A, MAMI) at the same kinematics. Nevertheless, the reported CLAS model uncertainties are nearly a factor of two smaller compared to the ones of MAMI and Hall-A. One can note here that in the CLAS analysis only two theoretical models have been utilized for the quantification of the model uncertainties, while for the MAMI and Hall-A data a more complete study that engages four theoretical models, as well as an experimental measurement of the photon excitation channel has been considered (as described in detail in earlier sections of this document). Taking the above considerations into account one may wonder if the small improvement that we observe in the $r_{\rm n}$ uncertainty by the inclusion of the CLAS data, in a region where $G_{\rm E}^{\rm n}$ data already exist, is affected by a potential underestimation of these uncertainties. In any case, and for the completeness of this work, we report the results for the $\langle r_{\rm n}^2 \rangle$ extraction when the CLAS data are included in the analysis, but we suggest that this result is considered with a level of reservation.


\section{Flavor dependent charge densities}

The neutron and the proton charge densities~\cite{cMiller:2007uy} can be extracted in the infinite-momentum-frame through: 
\begin{linenomath}\begin{equation}\label{density}
\rho(b)= \int_0^\infty\ {dQ\ Q\ \over  2 \pi}J_0(Q b) {G_{\rm E}(Q^2)+\tau G_{\rm M}(Q^2)\over 1+\tau}
\end{equation}\end{linenomath}
where $b$ is the transverse distance, $\tau={Q^2/4m^2}$ and $J_0$ a cylindrical Bessel function. From the neutron and proton densities, invoking charge symmetry, and neglecting the $s\bar{s}$ contribution, we then extract the $u$- and $d$-quark densities, where
\begin{linenomath}\begin{equation}\label{qdensit}
\rho_{\rm u}(b)=\rho_{\rm p}(b)+\rho_{\rm n}(b)/2 \qquad
\rho_{\rm d}(b)=\rho_{\rm p}(b)+2\rho_{\rm n}(b).
\end{equation}\end{linenomath}
Here we utilize the most recent parametrizations of the nucleon form factors
For the neutron electric form factor we use the parametrization determined in this work i.e. the one of Eq.~\ref{modgalster} (Table~\ref{radparam}).
For the two magnetic form factors,$G_{\rm M}^{\rm p}$ and $G_{\rm M}^{\rm n}$, we use~\cite{cYe:2017gyb}. For $G_{\rm E}^{\rm p}$ we have performed an updated parametrization for the following reasons: i) so that we may include important and recent measurements from~\cite{cXiong:2019umf} that were not yet available in~\cite{cYe:2017gyb}, and ii) so that we may eliminate a constraint on the proton charge radius that has been enforced in~\cite{cYe:2017gyb}, which is not in agreement with recent measurements of the proton charge radius. Here we have used a form that has been widely adopted in the past:
\begin{linenomath}\begin{equation}\label{gepform} 
G^{\rm p}_{\rm E}=\frac{1+(\sum\limits_{i=1}^{2} a_i x_i^i)}{(1 + \sum\limits_{j=1}^{4} b_j x_j^j)} 
\end{equation}\end{linenomath}
The $G^{\rm p}_{\rm E}$ fitted parameters that we have derived are given in Table~\ref{gepparam}.
\begin{table}[t]
\centering
\begin{tabular}{|l|l|l|l|l|l|}
\hline
$\alpha_{1}$ & $\alpha_{2}$ &$b_{1}$ &$b_{2}$ &$b_{3}$ &$b_{4}$ \\ \hline
 0.13830$\pm$0.12960& 0.00011$\pm$0.00093 &11.1944$\pm$0.14021 &19.6659$\pm$1.1020 &30.5455$\pm$ 5.45231 & 2.35740 $\pm$0.03289 \\
\hline
\end{tabular}
\caption{\label{gepparam} The $G^{\rm p}_{\rm E}$ fitted parameters.}
\end{table}

\begin{figure}[t!]
\centering
\includegraphics[width=12.0cm]{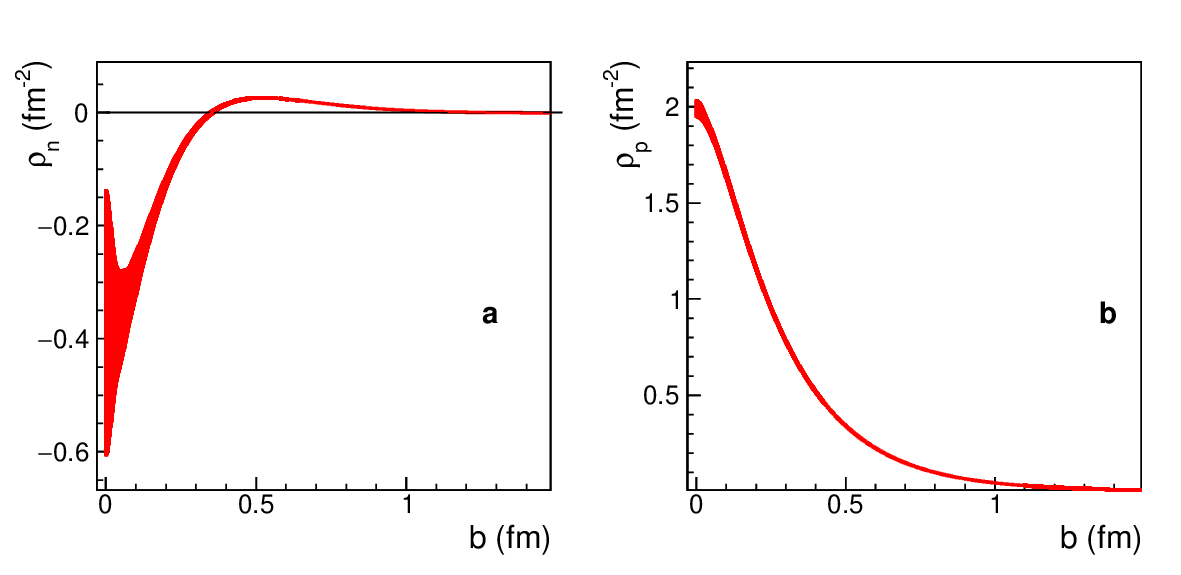}
\caption{The neutron charge density (panel a) and the proton charge density (panel b).}
\label{fig-densities}
\end{figure}

\begin{figure}[h!]
\centering
\includegraphics[width=12.0cm]{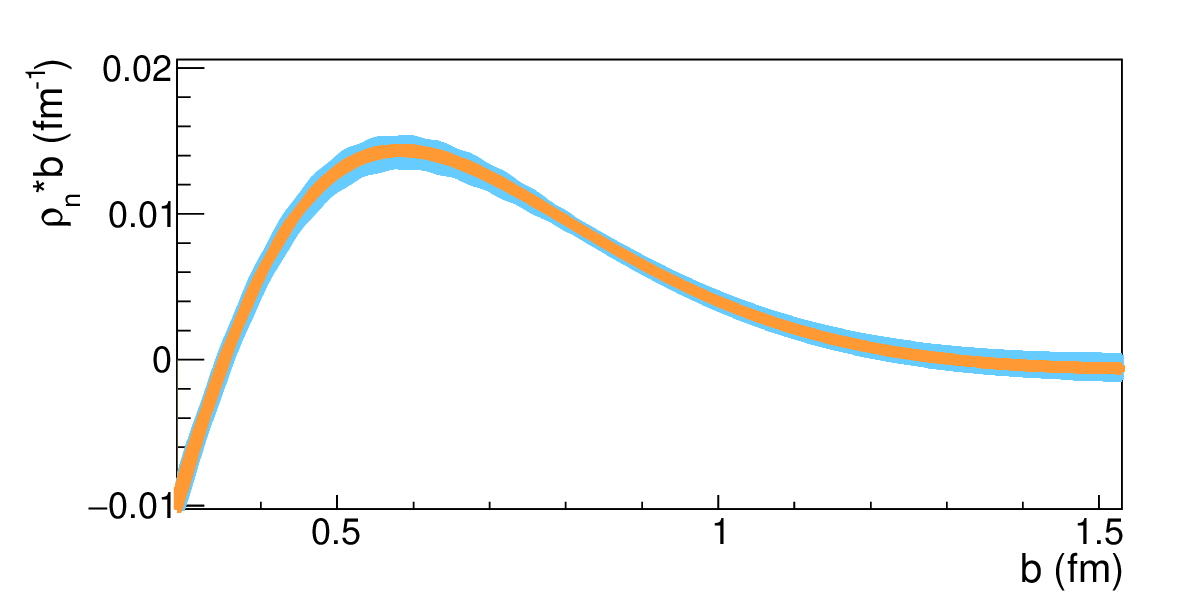}
\caption{The neutron charge density at it's surface, as derived with (without) our $G_{\rm E}^{\rm n}$ data is shown by the inner (outer) band.}
\label{neutronsurface}
\end{figure}

The extracted neutron ($\rho_{\rm n}$), proton ($\rho_{\rm p}$) are presented in Fig.~\ref{fig-densities}. The $u$- and $d$-quark densities with their experimental uncertainties are shown at Fig.~5c in the paper. The uncertainties in the charge density distributions result from the uncertainties of the nucleon form factor parametrizations, as they have been determined from the fits of the experimental data. Our data, being particularly sensitive to neutrons long distance structure, offer a factor of 2 improvement in the precision of the neutron charge density at its surface as seen in Fig.~\ref{neutronsurface}.

\section{Lattice QCD results}

To extract the lattice data presented in this work, we utilize results from a recent calculation by the Extended Twisted Mass Collaboration (ETMC) on the electromagnetic form factors of the proton and neutron~\cite{cAlexandrou:2018sjm}. The lattice calculation is pioneering in more than one ways. First, the numerical simulations have been performed taking into account the up, down, strange and charm quark in the sea ($N_{\rm f}=2+1+1$ of twisted mass clover-improved fermions). All of the quark masses are fixed to their physical value (physical point), reproducing a pion mass of 139 MeV. In addition, the ensemble has volume with spatial extent 5.12 fm, and a lattice spacing $a=0.08$ fm. Having a lattice spacing below 0.1 fm, and a large volume,  significantly suppressed finite-volume effects and discretizations effects. 

Beyond the state-of-the-art ensemble used for the calculation of the electromagnetic form factors, several sources of systematic uncertainties are controlled. Dominance of ground state is reliably established by three methods to analyze the data: single-state fit, two-state fit, and the so-called summation method~\cite{cMaiani:1987by,cCapitani:2012gj}. Furthermore, both the connected and disconnected contributions are included to extract the proton and neutron electromagnetic form factors at the physical point. The achieved accuracy for both connected and disconnected contributions is very high, which is a very challenging task for simulations at the physical point. In fact, it is also shown that the disconnected contributions are non-negligible, reaching up to 15$\%$. Along the same lines, the strange quark contribution to the electromagnetic form factors has been calculated~\cite{cAlexandrou:2019olr}, which is, however, beyond the scope of this work.

Here we used the aforementioned data to extract the ratio between the neutron electric and magnetic form factor ($G_{\rm E}^{\rm n}/G_{\rm M}^{\rm n}$). The results are shown in Table~\ref{tablatgegm}. The very good agreement of the results with the experimental data indicates that lattice QCD has advanced significantly, leading to results of high reliability. The errors include both statistical and systematic uncertainties added in quadrature. For the ratios we perform a jackknife statistical-error analysis, and an error propagation analysis using the individual data on $G_{\rm E}^{\rm n}$ and $G_{\rm M}^{\rm n}$. The resulting mean values and errors are almost identical between the two methods.

\begin{table}[h!]
\centering
\begin{tabular}{|l|l|l|}
\hline
$Q^2 ({\rm GeV}/c)^2$ & $\frac{G^{\rm n}_{\rm E}}{G^{\rm n}_{\rm M}}$ & $\delta(\frac{G^{\rm n}_{\rm E}}{G^{\rm n}_{\rm M}})$  \\ \hline
0.057 & 0.0116 & 0.0032 \\
0.113 & 0.0198 & 0.0040 \\
0.167 & 0.0298 & 0.0060 \\
0.219 & 0.0298 & 0.0077 \\
0.271 & 0.0427 & 0.0071 \\
0.321 & 0.0543 & 0.0084 \\
0.417 & 0.0618 & 0.0134 \\
0.464 & 0.0688 & 0.0137 \\
0.510 & 0.0639 & 0.0162 \\
0.554 & 0.0658 & 0.0176 \\
0.598 & 0.0797 & 0.0300 \\
0.642 & 0.0751 & 0.0228 \\
0.684 & 0.0889 & 0.0251 \\
0.767 & 0.1167 & 0.0715 \\
0.807 & 0.0702 & 0.0397 \\
0.847 & 0.0928 & 0.0403 \\
0.886 & 0.1392 & 0.0626 \\
0.925 & 0.1267 & 0.0638 \\
0.963 & 0.0916 & 0.0571 \\
\hline
\end{tabular}
\caption{\label{tablatgegm} Lattice results for $\frac{G^{\rm n}_{\rm E}}{G^{\rm n}_{\rm M}}$. }
\end{table}



\newpage

\bibliography{neutron_suppl_combine}

\begin{thebibliography}{10}
\urlstyle{rm}
\expandafter\ifx\csname url\endcsname\relax
  \def\url#1{\texttt{#1}}\fi
\expandafter\ifx\csname urlprefix\endcsname\relax\def\urlprefix{URL }\fi
\expandafter\ifx\csname doiprefix\endcsname\relax\def\doiprefix{DOI: }\fi
\providecommand{\bibinfo}[2]{#2}
\providecommand{\eprint}[2][]{\url{#2}}

\bibitem{cBlomberg:2015zma}
\bibinfo{author}{Blomberg, A.} \emph{et~al.}
\newblock \bibinfo{journal}{\bibinfo{title}{{Electroexcitation of the
  $\Delta^{+}(1232)$ at low momentum transfer}}}.
\newblock {\emph{\JournalTitle{Phys. Lett.}}} \textbf{\bibinfo{volume}{B760}},
  \bibinfo{pages}{267--272} (\bibinfo{year}{2016}).

\bibitem{cStave:2006ea}
\bibinfo{author}{Stave, S.} \emph{et~al.}
\newblock \bibinfo{journal}{\bibinfo{title}{{Lowest $Q^2$ Measurement of the
  $\gamma*$ p ---> $\Delta$ Reaction: Probing the Pionic Contribution}}}.
\newblock {\emph{\JournalTitle{Eur. Phys. J.}}} \textbf{\bibinfo{volume}{A30}},
  \bibinfo{pages}{471--476} (\bibinfo{year}{2006}).

\bibitem{cSparveris:2013ena}
\bibinfo{author}{Sparveris, N.} \emph{et~al.}
\newblock \bibinfo{journal}{\bibinfo{title}{{Measurements of the $\gamma$*p
  $\rightarrow$ $\Delta$ reaction at low $Q^{2}$}}}.
\newblock {\emph{\JournalTitle{Eur. Phys. J.}}} \textbf{\bibinfo{volume}{A49}},
  \bibinfo{pages}{136} (\bibinfo{year}{2013}).

\bibitem{cSparveris:2006uk}
\bibinfo{author}{Sparveris, N.~F.} \emph{et~al.}
\newblock \bibinfo{journal}{\bibinfo{title}{{Determination of quadrupole
  strengths in the $\gamma* p \rightarrow \Delta(1232)$ transition at $Q^2$ =
  0.20 $(GeV/c)^2$}}}.
\newblock {\emph{\JournalTitle{Phys. Lett.}}} \textbf{\bibinfo{volume}{B651}},
  \bibinfo{pages}{102--107} (\bibinfo{year}{2007}).

\bibitem{cBlomberg:2019caf}
\bibinfo{author}{Blomberg, A.} \emph{et~al.}
\newblock \bibinfo{journal}{\bibinfo{title}{{Virtual Compton Scattering
  measurements in the nucleon resonance region}}}.
\newblock {\emph{\JournalTitle{Eur. Phys. J. A}}}
  \textbf{\bibinfo{volume}{55}}, \bibinfo{pages}{182} (\bibinfo{year}{2019}).
\newblock \eprint{1901.08951}.

\bibitem{cAznauryan:2009mx}
\bibinfo{author}{Aznauryan, I.~G.} \emph{et~al.}
\newblock \bibinfo{journal}{\bibinfo{title}{{Electroexcitation of nucleon
  resonances from CLAS data on single pion electroproduction}}}.
\newblock {\emph{\JournalTitle{Phys. Rev.}}} \textbf{\bibinfo{volume}{C80}},
  \bibinfo{pages}{055203} (\bibinfo{year}{2009}).

\bibitem{cKamalov:1999hs}
\bibinfo{author}{Kamalov, S.} \& \bibinfo{author}{Yang, S.~N.}
\newblock \bibinfo{journal}{\bibinfo{title}{{Pion cloud and the $Q^2$
  dependence of $\gamma*$ N <---> $\Delta$ transition form-factors}}}.
\newblock {\emph{\JournalTitle{Phys. Rev. Lett.}}}
  \textbf{\bibinfo{volume}{83}}, \bibinfo{pages}{4494--4497}
  (\bibinfo{year}{1999}).

\bibitem{cSato:2000jf}
\bibinfo{author}{Sato, T.} \& \bibinfo{author}{Lee, T.}
\newblock \bibinfo{journal}{\bibinfo{title}{{Dynamical study of the Delta
  excitation in N (e, e-prime pi) reactions}}}.
\newblock {\emph{\JournalTitle{Phys. Rev. C}}} \textbf{\bibinfo{volume}{63}},
  \bibinfo{pages}{055201} (\bibinfo{year}{2001}).

\bibitem{cDrechsel:1998hk}
\bibinfo{author}{Drechsel, D.}, \bibinfo{author}{Hanstein, O.},
  \bibinfo{author}{Kamalov, S.~S.} \& \bibinfo{author}{Tiator, L.}
\newblock \bibinfo{journal}{\bibinfo{title}{{A Unitary isobar model for pion
  photoproduction and electroproduction on the proton up to 1-GeV}}}.
\newblock {\emph{\JournalTitle{Nucl. Phys.}}} \textbf{\bibinfo{volume}{A645}},
  \bibinfo{pages}{145--174} (\bibinfo{year}{1999}).

\bibitem{cKamalov:2001qg}
\bibinfo{author}{Kamalov, S.}, \bibinfo{author}{Chen, G.-Y.},
  \bibinfo{author}{Yang, S.-N.}, \bibinfo{author}{Drechsel, D.} \&
  \bibinfo{author}{Tiator, L.}
\newblock \bibinfo{journal}{\bibinfo{title}{{Pi0 photoproduction and
  electroproduction at threshold within a dynamical model}}}.
\newblock {\emph{\JournalTitle{Phys. Lett. B}}} \textbf{\bibinfo{volume}{522}},
  \bibinfo{pages}{27--36} (\bibinfo{year}{2001}).

\bibitem{cArndt:2002xv}
\bibinfo{author}{Arndt, R.~A.}, \bibinfo{author}{Briscoe, W.~J.},
  \bibinfo{author}{Strakovsky, I.~I.} \& \bibinfo{author}{Workman, R.~L.}
\newblock \bibinfo{journal}{\bibinfo{title}{{Analysis of pion photoproduction
  data}}}.
\newblock {\emph{\JournalTitle{Phys. Rev.}}} \textbf{\bibinfo{volume}{C66}},
  \bibinfo{pages}{055213} (\bibinfo{year}{2002}).

\bibitem{cBuchmann:2004ia}
\bibinfo{author}{Buchmann, A.~J.}
\newblock \bibinfo{journal}{\bibinfo{title}{{Electromagnetic N ---> $\Delta$
  transition and neutron form-factors}}}.
\newblock {\emph{\JournalTitle{Phys. Rev. Lett.}}}
  \textbf{\bibinfo{volume}{93}}, \bibinfo{pages}{212301}
  (\bibinfo{year}{2004}).

\bibitem{cVanderhaeghen:2007}
\bibinfo{author}{Pascalutsa, V.} \& \bibinfo{author}{Vanderhaeghen, M.}
\newblock \bibinfo{journal}{\bibinfo{title}{{Large-$N_c$ relations for the
  electromagnetic nucleon-to-$\Delta$ form factors}}}.
\newblock {\emph{\JournalTitle{Phys. Rev.}}} \textbf{\bibinfo{volume}{D76}},
  \bibinfo{pages}{111501(R)} (\bibinfo{year}{2007}).

\bibitem{cRiordan:2010id}
\bibinfo{author}{Riordan, S.} \emph{et~al.}
\newblock \bibinfo{journal}{\bibinfo{title}{{Measurements of the Electric Form
  Factor of the Neutron up to $Q^2$=3.4 $GeV^2$ using the Reaction
  $\vec{^3He}(\vec{e},e'n)pp$}}}.
\newblock {\emph{\JournalTitle{Phys. Rev. Lett.}}}
  \textbf{\bibinfo{volume}{105}}, \bibinfo{pages}{262302}
  (\bibinfo{year}{2010}).

\bibitem{cGeis:2008aa}
\bibinfo{author}{Geis, E.} \emph{et~al.}
\newblock \bibinfo{journal}{\bibinfo{title}{{The Charge Form Factor of the
  Neutron at Low Momentum Transfer from the $\vec{^2H} (\vec{e}, e^{\prime} n)
  p$ Reaction}}}.
\newblock {\emph{\JournalTitle{Phys. Rev. Lett.}}}
  \textbf{\bibinfo{volume}{101}}, \bibinfo{pages}{042501}
  (\bibinfo{year}{2008}).

\bibitem{cYe:2017gyb}
\bibinfo{author}{Ye, Z.}, \bibinfo{author}{Arrington, J.},
  \bibinfo{author}{Hill, R.~J.} \& \bibinfo{author}{Lee, G.}
\newblock \bibinfo{journal}{\bibinfo{title}{{Proton and Neutron Electromagnetic
  Form Factors and Uncertainties}}}.
\newblock {\emph{\JournalTitle{Phys. Lett.}}} \textbf{\bibinfo{volume}{B777}},
  \bibinfo{pages}{8--15} (\bibinfo{year}{2018}).

\bibitem{cKelly:2004hm}
\bibinfo{author}{Kelly, J.~J.}
\newblock \bibinfo{journal}{\bibinfo{title}{{Simple parametrization of nucleon
  form factors}}}.
\newblock {\emph{\JournalTitle{Phys. Rev.}}} \textbf{\bibinfo{volume}{C70}},
  \bibinfo{pages}{068202} (\bibinfo{year}{2004}).

\bibitem{cSparveris:2004jn}
\bibinfo{author}{Sparveris, N.~F.} \emph{et~al.}
\newblock \bibinfo{journal}{\bibinfo{title}{{Investigation of the conjectured
  nucleon deformation at low momentum transfer}}}.
\newblock {\emph{\JournalTitle{Phys. Rev. Lett.}}}
  \textbf{\bibinfo{volume}{94}}, \bibinfo{pages}{022003}
  (\bibinfo{year}{2005}).

\bibitem{cElsner:2005cz}
\bibinfo{author}{Elsner, D.} \emph{et~al.}
\newblock \bibinfo{journal}{\bibinfo{title}{{Measurement of the LT-asymmetry in
  $\pi^{0}$ electroproduction at the energy of the $\Delta(1232)$ resonance}}}.
\newblock {\emph{\JournalTitle{Eur. Phys. J.}}} \textbf{\bibinfo{volume}{A27}},
  \bibinfo{pages}{91--97} (\bibinfo{year}{2006}).

\bibitem{cKelly:2005jy}
\bibinfo{author}{Kelly, J.~J.} \emph{et~al.}
\newblock \bibinfo{journal}{\bibinfo{title}{{Recoil polarization measurements
  for neutral pion electroproduction at $Q^2$ = 1 $(GeV/c)^2$ near the Delta
  resonance}}}.
\newblock {\emph{\JournalTitle{Phys. Rev.}}} \textbf{\bibinfo{volume}{C75}},
  \bibinfo{pages}{025201} (\bibinfo{year}{2007}).

\bibitem{cMadey:2003av}
\bibinfo{author}{Madey, R.} \emph{et~al.}
\newblock \bibinfo{journal}{\bibinfo{title}{{Measurements of $G_{E}^n /
  G_{M}^n$ from the $^2H(\vec{e},e^{\prime} \vec{n})^1H$ reaction to $Q^2$ =
  1.45 $(GeV/c)^2$}}}.
\newblock {\emph{\JournalTitle{Phys. Rev. Lett.}}}
  \textbf{\bibinfo{volume}{91}}, \bibinfo{pages}{122002}
  (\bibinfo{year}{2003}).

\bibitem{cSchlimme:2013eoz}
\bibinfo{author}{Schlimme, B.~S.} \emph{et~al.}
\newblock \bibinfo{journal}{\bibinfo{title}{{Measurement of the neutron
  electric to magnetic form factor ratio at $Q^2$ = 1.58 $GeV^2$ using the
  reaction $^3\vec{He}(\vec{e},e'n)pp$}}}.
\newblock {\emph{\JournalTitle{Phys. Rev. Lett.}}}
  \textbf{\bibinfo{volume}{111}}, \bibinfo{pages}{132504}
  (\bibinfo{year}{2013}).

\bibitem{cGlazier:2004ny}
\bibinfo{author}{Glazier, D.~I.} \emph{et~al.}
\newblock \bibinfo{journal}{\bibinfo{title}{{Measurement of the electric
  form-factor of the neutron at $Q^2$ = 0.3 $(GeV/c)^2$ to 0.8 $(GeV/c)^2$}}}.
\newblock {\emph{\JournalTitle{Eur. Phys. J.}}} \textbf{\bibinfo{volume}{A24}},
  \bibinfo{pages}{101--109} (\bibinfo{year}{2005}).

\bibitem{cPlaster:2005cx}
\bibinfo{author}{Plaster, B.} \emph{et~al.}
\newblock \bibinfo{journal}{\bibinfo{title}{{Measurements of the neutron
  electric to magnetic form-factor ratio $G_{En} / G_{Mn}$ via the
  $^2H(\vec{e},e^{\prime},\vec{n})^1H$ reaction to $Q^2$ =1.45 $(GeV/c)^2$}}}.
\newblock {\emph{\JournalTitle{Phys. Rev.}}} \textbf{\bibinfo{volume}{C73}},
  \bibinfo{pages}{025205} (\bibinfo{year}{2006}).

\bibitem{cZhu:2001md}
\bibinfo{author}{Zhu, H.} \emph{et~al.}
\newblock \bibinfo{journal}{\bibinfo{title}{{A Measurement of the electric
  form-factor of the neutron through $\vec{d} (\vec{e}, e^{\prime} n)p$ at
  $Q^2$ = 0.5 $(GeV/c)^2$}}}.
\newblock {\emph{\JournalTitle{Phys. Rev. Lett.}}}
  \textbf{\bibinfo{volume}{87}}, \bibinfo{pages}{081801}
  (\bibinfo{year}{2001}).

\bibitem{cWarren:2003ma}
\bibinfo{author}{Warren, G.} \emph{et~al.}
\newblock \bibinfo{journal}{\bibinfo{title}{{Measurement of the electric
  form-factor of the neutron at $Q^2$ = 0.5 and 1.0 $GeV^2/c^2$}}}.
\newblock {\emph{\JournalTitle{Phys. Rev. Lett.}}}
  \textbf{\bibinfo{volume}{92}}, \bibinfo{pages}{042301}
  (\bibinfo{year}{2004}).

\bibitem{cRohe:1999sh}
\bibinfo{author}{Rohe, D.} \emph{et~al.}
\newblock \bibinfo{journal}{\bibinfo{title}{{Measurement of the neutron
  electric form-factor $G_{(en)}$ at $0.67 (GeV/c)^2$ via
  $\vec{^{3}He}(\vec{e},e' n)$}}}.
\newblock {\emph{\JournalTitle{Phys. Rev. Lett.}}}
  \textbf{\bibinfo{volume}{83}}, \bibinfo{pages}{4257--4260}
  (\bibinfo{year}{1999}).

\bibitem{cBermuth:2003qh}
\bibinfo{author}{Bermuth, J.} \emph{et~al.}
\newblock \bibinfo{journal}{\bibinfo{title}{{The Neutron charge form-factor and
  target analyzing powers from $\vec{^3He} (\vec{e},e^{\prime} n)$
  scattering}}}.
\newblock {\emph{\JournalTitle{Phys. Lett.}}} \textbf{\bibinfo{volume}{B564}},
  \bibinfo{pages}{199--204} (\bibinfo{year}{2003}).

\bibitem{cEden.50.R1749}
\bibinfo{author}{Eden, T.} \emph{et~al.}
\newblock \bibinfo{journal}{\bibinfo{title}{{Electric form factor of the
  neutron from the $^{2}\vec{H}$($\vec{e}$,$e^{\prime}$n)$^{1}H$ reaction at
  $Q^{2}$=0.255 $(GeV/c)^{2}$}}}.
\newblock {\emph{\JournalTitle{Phys. Rev. C}}} \textbf{\bibinfo{volume}{50}},
  \bibinfo{pages}{R1749--R1753} (\bibinfo{year}{1994}).

\bibitem{costrick.83.276}
\bibinfo{author}{Ostrick, M.} \emph{et~al.}
\newblock \bibinfo{journal}{\bibinfo{title}{{Measurement of the Neutron
  Electric Form Factor ${G}_{E,n}$ in the Quasifree
  $^{2}H(\stackrel{\ensuremath{\rightarrow}}{\mathit{e}},{\mathit{e}}^{\ensuremath{'}}\stackrel{\ensuremath{\rightarrow}}{\mathit{n}})\mathit{p}$
  Reaction}}}.
\newblock {\emph{\JournalTitle{Phys. Rev. Lett.}}}
  \textbf{\bibinfo{volume}{83}}, \bibinfo{pages}{276--279}
  (\bibinfo{year}{1999}).

\bibitem{cgolak.63.034006}
\bibinfo{author}{Golak, J.}, \bibinfo{author}{Ziemer, G.},
  \bibinfo{author}{Kamada, H.}, \bibinfo{author}{Wita\l{}a, H.} \&
  \bibinfo{author}{Gl\"ockle, W.}
\newblock \bibinfo{journal}{\bibinfo{title}{Extraction of electromagnetic
  neutron form factors through inclusive and exclusive polarized electron
  scattering on a polarized ${}^{3}\mathrm{He}$ target}}.
\newblock {\emph{\JournalTitle{Phys. Rev. C}}} \textbf{\bibinfo{volume}{63}},
  \bibinfo{pages}{034006} (\bibinfo{year}{2001}).

\bibitem{cHerberg:1999ud}
\bibinfo{author}{Herberg, C.} \emph{et~al.}
\newblock \bibinfo{journal}{\bibinfo{title}{{Determination of the neutron
  electric form-factor in the D(e,e' n)p reaction and the influence of nuclear
  binding}}}.
\newblock {\emph{\JournalTitle{Eur. Phys. J.}}} \textbf{\bibinfo{volume}{A5}},
  \bibinfo{pages}{131--135} (\bibinfo{year}{1999}).

\bibitem{cGalster:1971kv}
\bibinfo{author}{Galster, S.} \emph{et~al.}
\newblock \bibinfo{journal}{\bibinfo{title}{{Elastic electron-deuteron
  scattering and the electric neutron form factor at four-momentum transfers
  5fm$^{-2} < q^2 < 14$fm$^{-2}$}}}.
\newblock {\emph{\JournalTitle{Nucl. Phys. B}}} \textbf{\bibinfo{volume}{32}},
  \bibinfo{pages}{221--237} (\bibinfo{year}{1971}).

\bibitem{gentile:2011}
\bibinfo{author}{Gentile, T.} \& \bibinfo{author}{Crawford, C.}
\newblock \bibinfo{journal}{\bibinfo{title}{{Neutron charge radius and the
  neutron electric form factor}}}.
\newblock {\emph{\JournalTitle{Phys. Rev. C}}} \textbf{\bibinfo{volume}{83}},
  \bibinfo{pages}{055203} (\bibinfo{year}{2011}).

\bibitem{cKopecky:1997rw}
\bibinfo{author}{Kopecky, S.} \emph{et~al.}
\newblock \bibinfo{journal}{\bibinfo{title}{{Neutron charge radius determined
  from the energy dependence of the neutron transmission of liquid Pb-208 and
  Bi-209}}}.
\newblock {\emph{\JournalTitle{Phys. Rev.}}} \textbf{\bibinfo{volume}{C56}},
  \bibinfo{pages}{2229--2237} (\bibinfo{year}{1997}).

\bibitem{cKoester:1995nx}
\bibinfo{author}{Koester, L.} \emph{et~al.}
\newblock \bibinfo{journal}{\bibinfo{title}{{Neutron electron scattering length
  and electric polarizability of the neutron derived from cross-sections of
  bismuth and of lead and its isotopes}}}.
\newblock {\emph{\JournalTitle{Phys. Rev.}}} \textbf{\bibinfo{volume}{C51}},
  \bibinfo{pages}{3363--3371} (\bibinfo{year}{1995}).

\bibitem{cAleksandrov:1986mw}
\bibinfo{author}{Aleksandrov, {\relax Yu}.~A.}, \bibinfo{author}{Vrana, M.},
  \bibinfo{author}{Manrique, G.~J.}, \bibinfo{author}{Machekhina, T.~A.} \&
  \bibinfo{author}{Sedlakova, L.~N.}
\newblock \bibinfo{journal}{\bibinfo{title}{{Neutron rms radius and electric
  polarizability from data on the interaction of slow neutrons with bismuth}}}.
\newblock {\emph{\JournalTitle{Sov. J. Nucl. Phys.}}}
  \textbf{\bibinfo{volume}{44}}, \bibinfo{pages}{900--902}
  (\bibinfo{year}{1986}).

\bibitem{cKrohn:1973re}
\bibinfo{author}{Krohn, V.~E.} \& \bibinfo{author}{Ringo, G.~R.}
\newblock \bibinfo{journal}{\bibinfo{title}{{Reconsiderations of the electron -
  neutron scattering length as measured by the scattering of thermal neutrons
  by noble gases}}}.
\newblock {\emph{\JournalTitle{Phys. Rev.}}} \textbf{\bibinfo{volume}{D8}},
  \bibinfo{pages}{1305--1307} (\bibinfo{year}{1973}).

\bibitem{cMiller:2007uy}
\bibinfo{author}{Miller, G.~A.}
\newblock \bibinfo{journal}{\bibinfo{title}{{Charge Density of the Neutron and
  Proton }}}.
\newblock {\emph{\JournalTitle{Phys. Rev. Lett.}}}
  \textbf{\bibinfo{volume}{99}}, \bibinfo{pages}{112001}
  (\bibinfo{year}{2007}).

\bibitem{cXiong:2019umf}
\bibinfo{author}{Xiong, W.} \emph{et~al.}
\newblock \bibinfo{journal}{\bibinfo{title}{{A small proton charge radius from
  an electron–proton scattering experiment}}}.
\newblock {\emph{\JournalTitle{Nature}}} \textbf{\bibinfo{volume}{575}},
  \bibinfo{pages}{147--150} (\bibinfo{year}{2019}).

\bibitem{cAlexandrou:2018sjm}
\bibinfo{author}{Alexandrou, C.} \emph{et~al.}
\newblock \bibinfo{journal}{\bibinfo{title}{{Proton and neutron electromagnetic
  form factors from lattice QCD}}}.
\newblock {\emph{\JournalTitle{Phys. Rev.}}} \textbf{\bibinfo{volume}{D100}},
  \bibinfo{pages}{014509} (\bibinfo{year}{2019}).

\bibitem{cMaiani:1987by}
\bibinfo{author}{Maiani, L.}, \bibinfo{author}{Martinelli, G.},
  \bibinfo{author}{Paciello, M.} \& \bibinfo{author}{Taglienti, B.}
\newblock \bibinfo{journal}{\bibinfo{title}{{Scalar Densities and Baryon Mass
  Differences in Lattice {QCD} With Wilson Fermions}}}.
\newblock {\emph{\JournalTitle{Nucl. Phys. B}}} \textbf{\bibinfo{volume}{293}},
  \bibinfo{pages}{420}, \doiprefix\url{10.1016/0550-3213(87)90078-2}
  (\bibinfo{year}{1987}).

\bibitem{cCapitani:2012gj}
\bibinfo{author}{Capitani, S.} \emph{et~al.}
\newblock \bibinfo{journal}{\bibinfo{title}{{The nucleon axial charge from
  lattice QCD with controlled errors}}}.
\newblock {\emph{\JournalTitle{Phys. Rev. D}}} \textbf{\bibinfo{volume}{86}},
  \bibinfo{pages}{074502}, \doiprefix\url{10.1103/PhysRevD.86.074502}
  (\bibinfo{year}{2012}).
\newblock \eprint{1205.0180}.

\bibitem{cAlexandrou:2019olr}
\bibinfo{author}{Alexandrou, C.} \emph{et~al.}
\newblock \bibinfo{journal}{\bibinfo{title}{{Nucleon strange electromagnetic
  form factors}}}.
\newblock {\emph{\JournalTitle{Phys. Rev. D}}} \textbf{\bibinfo{volume}{101}},
  \bibinfo{pages}{031501}, \doiprefix\url{10.1103/PhysRevD.101.031501}
  (\bibinfo{year}{2020}).
\newblock \eprint{1909.10744}.

\end{thebibliography}


\begin{thebibliography}{10}
\urlstyle{rm}
\expandafter\ifx\csname url\endcsname\relax
  \def\url#1{\texttt{#1}}\fi
\expandafter\ifx\csname urlprefix\endcsname\relax\def\urlprefix{URL }\fi
\expandafter\ifx\csname doiprefix\endcsname\relax\def\doiprefix{DOI: }\fi
\providecommand{\bibinfo}[2]{#2}
\providecommand{\eprint}[2][]{\url{#2}}

\bibitem{pohl:2010}
\bibinfo{author}{Pohl, R.} \emph{et~al.}
\newblock \bibinfo{journal}{\bibinfo{title}{{The size of the proton.}}}
\newblock {\emph{\JournalTitle{Nature}}} \textbf{\bibinfo{volume}{466}},
  \bibinfo{pages}{213--216} (\bibinfo{year}{2010}).

\bibitem{protonpuzz:2013}
\bibinfo{author}{R.~Pohl, G.~A.~M., R.~Gilman} \& \bibinfo{author}{Pachucki,
  K.}
\newblock \bibinfo{journal}{\bibinfo{title}{{Muonic hydrogen and the proton
  radius puzzle.}}}
\newblock {\emph{\JournalTitle{Ann. Rev. Nucl. Part. Sci.}}}
  \textbf{\bibinfo{volume}{63}}, \bibinfo{pages}{175--204}
  (\bibinfo{year}{2013}).

\bibitem{Kopecky:1997rw}
\bibinfo{author}{Kopecky, S.} \emph{et~al.}
\newblock \bibinfo{journal}{\bibinfo{title}{{Neutron charge radius determined
  from the energy dependence of the neutron transmission of liquid Pb-208 and
  Bi-209}}}.
\newblock {\emph{\JournalTitle{Phys. Rev.}}} \textbf{\bibinfo{volume}{C56}},
  \bibinfo{pages}{2229--2237} (\bibinfo{year}{1997}).

\bibitem{Koester:1995nx}
\bibinfo{author}{Koester, L.} \emph{et~al.}
\newblock \bibinfo{journal}{\bibinfo{title}{{Neutron electron scattering length
  and electric polarizability of the neutron derived from cross-sections of
  bismuth and of lead and its isotopes}}}.
\newblock {\emph{\JournalTitle{Phys. Rev.}}} \textbf{\bibinfo{volume}{C51}},
  \bibinfo{pages}{3363--3371} (\bibinfo{year}{1995}).

\bibitem{Aleksandrov:1986mw}
\bibinfo{author}{Aleksandrov, {\relax Yu}.~A.}, \bibinfo{author}{Vrana, M.},
  \bibinfo{author}{Manrique, G.~J.}, \bibinfo{author}{Machekhina, T.~A.} \&
  \bibinfo{author}{Sedlakova, L.~N.}
\newblock \bibinfo{journal}{\bibinfo{title}{{Neutron rms radius and electric
  polarizability from data on the interaction of slow neutrons with bismuth}}}.
\newblock {\emph{\JournalTitle{Sov. J. Nucl. Phys.}}}
  \textbf{\bibinfo{volume}{44}}, \bibinfo{pages}{900--902}
  (\bibinfo{year}{1986}).

\bibitem{Krohn:1973re}
\bibinfo{author}{Krohn, V.~E.} \& \bibinfo{author}{Ringo, G.~R.}
\newblock \bibinfo{journal}{\bibinfo{title}{{Reconsiderations of the electron -
  neutron scattering length as measured by the scattering of thermal neutrons
  by noble gases}}}.
\newblock {\emph{\JournalTitle{Phys. Rev.}}} \textbf{\bibinfo{volume}{D8}},
  \bibinfo{pages}{1305--1307} (\bibinfo{year}{1973}).

\bibitem{Madey:2003av}
\bibinfo{author}{Madey, R.} \emph{et~al.}
\newblock \bibinfo{journal}{\bibinfo{title}{{Measurements of $G_{E}^{\rm n} /
  G_{M}^{\rm n}$ from the $^2H(\vec{e},e^{\prime} \vec{n})^1H$ reaction to $Q^2$ =
  1.45 $({\rm GeV}/c)^2$}}}.
\newblock {\emph{\JournalTitle{Phys. Rev. Lett.}}}
  \textbf{\bibinfo{volume}{91}}, \bibinfo{pages}{122002}
  (\bibinfo{year}{2003}).

\bibitem{Schlimme:2013eoz}
\bibinfo{author}{Schlimme, B.~S.} \emph{et~al.}
\newblock \bibinfo{journal}{\bibinfo{title}{{Measurement of the neutron
  electric to magnetic form factor ratio at $Q^2$ = 1.58 ${\rm GeV}^2$ using the
  reaction $^3\vec{He}(\vec{e},e'n)pp$}}}.
\newblock {\emph{\JournalTitle{Phys. Rev. Lett.}}}
  \textbf{\bibinfo{volume}{111}}, \bibinfo{pages}{132504}
  (\bibinfo{year}{2013}).

\bibitem{Riordan:2010id}
\bibinfo{author}{Riordan, S.} \emph{et~al.}
\newblock \bibinfo{journal}{\bibinfo{title}{{Measurements of the Electric Form
  Factor of the Neutron up to $Q^2$=3.4 ${\rm GeV}^2$ using the Reaction
  $\vec{^3He}(\vec{e},e'n)pp$}}}.
\newblock {\emph{\JournalTitle{Phys. Rev. Lett.}}}
  \textbf{\bibinfo{volume}{105}}, \bibinfo{pages}{262302}
  (\bibinfo{year}{2010}).

\bibitem{Glazier:2004ny}
\bibinfo{author}{Glazier, D.~I.} \emph{et~al.}
\newblock \bibinfo{journal}{\bibinfo{title}{{Measurement of the electric
  form-factor of the neutron at $Q^2$ = 0.3 $({\rm GeV}/c)^2$ to 0.8 $({\rm GeV}/c)^2$}}}.
\newblock {\emph{\JournalTitle{Eur. Phys. J.}}} \textbf{\bibinfo{volume}{A24}},
  \bibinfo{pages}{101--109} (\bibinfo{year}{2005}).

\bibitem{Plaster:2005cx}
\bibinfo{author}{Plaster, B.} \emph{et~al.}
\newblock \bibinfo{journal}{\bibinfo{title}{{Measurements of the neutron
  electric to magnetic form-factor ratio $G_{En} / G_{Mn}$ via the
  $^2H(\vec{e},e^{\prime},\vec{n})^1H$ reaction to $Q^2$ =1.45 $({\rm GeV}/c)^2$}}}.
\newblock {\emph{\JournalTitle{Phys. Rev.}}} \textbf{\bibinfo{volume}{C73}},
  \bibinfo{pages}{025205} (\bibinfo{year}{2006}).

\bibitem{Zhu:2001md}
\bibinfo{author}{Zhu, H.} \emph{et~al.}
\newblock \bibinfo{journal}{\bibinfo{title}{{A Measurement of the electric
  form-factor of the neutron through $\vec{d} (\vec{e}, e^{\prime} n)p$ at
  $Q^2$ = 0.5 $({\rm GeV}/c)^2$}}}.
\newblock {\emph{\JournalTitle{Phys. Rev. Lett.}}}
  \textbf{\bibinfo{volume}{87}}, \bibinfo{pages}{081801}
  (\bibinfo{year}{2001}).

\bibitem{Warren:2003ma}
\bibinfo{author}{Warren, G.} \emph{et~al.}
\newblock \bibinfo{journal}{\bibinfo{title}{{Measurement of the electric
  form-factor of the neutron at $Q^2$ = 0.5 and 1.0 ${\rm GeV}^2/c^2$}}}.
\newblock {\emph{\JournalTitle{Phys. Rev. Lett.}}}
  \textbf{\bibinfo{volume}{92}}, \bibinfo{pages}{042301}
  (\bibinfo{year}{2004}).

\bibitem{Rohe:1999sh}
\bibinfo{author}{Rohe, D.} \emph{et~al.}
\newblock \bibinfo{journal}{\bibinfo{title}{{Measurement of the neutron
  electric form-factor $G_{(en)}$ at $0.67 ({\rm GeV}/c)^2$ via
  $\vec{^{3}He}(\vec{e},e' n)$}}}.
\newblock {\emph{\JournalTitle{Phys. Rev. Lett.}}}
  \textbf{\bibinfo{volume}{83}}, \bibinfo{pages}{4257--4260}
  (\bibinfo{year}{1999}).

\bibitem{Passchier:1999ju}
\bibinfo{author}{Passchier, I.} \emph{et~al.}
\newblock \bibinfo{journal}{\bibinfo{title}{{The Charge form-factor of the
  neutron from the reaction polarized $^2H (\vec{e},e, e^{\prime} n)p$}}}.
\newblock {\emph{\JournalTitle{Nucl. Phys. A}}} \textbf{\bibinfo{volume}{663}},
  \bibinfo{pages}{421--424} (\bibinfo{year}{2000}).

\bibitem{Bermuth:2003qh}
\bibinfo{author}{Bermuth, J.} \emph{et~al.}
\newblock \bibinfo{journal}{\bibinfo{title}{{The Neutron charge form-factor and
  target analyzing powers from $\vec{^3He} (\vec{e},e^{\prime} n)$
  scattering}}}.
\newblock {\emph{\JournalTitle{Phys. Lett.}}} \textbf{\bibinfo{volume}{B564}},
  \bibinfo{pages}{199--204} (\bibinfo{year}{2003}).

\bibitem{Geis:2008aa}
\bibinfo{author}{Geis, E.} \emph{et~al.}
\newblock \bibinfo{journal}{\bibinfo{title}{{The Charge Form Factor of the
  Neutron at Low Momentum Transfer from the $\vec{^2H} (\vec{e}, e^{\prime} n)
  p$ Reaction}}}.
\newblock {\emph{\JournalTitle{Phys. Rev. Lett.}}}
  \textbf{\bibinfo{volume}{101}}, \bibinfo{pages}{042501}
  (\bibinfo{year}{2008}).

\bibitem{Eden.50.R1749}
\bibinfo{author}{Eden, T.} \emph{et~al.}
\newblock \bibinfo{journal}{\bibinfo{title}{{Electric form factor of the
  neutron from the $^{2}\vec{H}$($\vec{e}$,$e^{\prime}$n)$^{1}H$ reaction at
  $Q^{2}$=0.255 $({\rm GeV}/c)^{2}$}}}.
\newblock {\emph{\JournalTitle{Phys. Rev. C}}} \textbf{\bibinfo{volume}{50}},
  \bibinfo{pages}{R1749--R1753} (\bibinfo{year}{1994}).

\bibitem{ostrick.83.276}
\bibinfo{author}{Ostrick, M.} \emph{et~al.}
\newblock \bibinfo{journal}{\bibinfo{title}{{Measurement of the Neutron
  Electric Form Factor ${G}_{E,n}$ in the Quasifree
  $^{2}H(\stackrel{\ensuremath{\rightarrow}}{\mathit{e}},{\mathit{e}}^{\ensuremath{'}}\stackrel{\ensuremath{\rightarrow}}{\mathit{n}})\mathit{p}$
  Reaction}}}.
\newblock {\emph{\JournalTitle{Phys. Rev. Lett.}}}
  \textbf{\bibinfo{volume}{83}}, \bibinfo{pages}{276--279}
  (\bibinfo{year}{1999}).

\bibitem{golak.63.034006}
\bibinfo{author}{Golak, J.}, \bibinfo{author}{Ziemer, G.},
  \bibinfo{author}{Kamada, H.}, \bibinfo{author}{Wita\l{}a, H.} \&
  \bibinfo{author}{Gl\"ockle, W.}
\newblock \bibinfo{journal}{\bibinfo{title}{Extraction of electromagnetic
  neutron form factors through inclusive and exclusive polarized electron
  scattering on a polarized ${}^{3}\mathrm{He}$ target}}.
\newblock {\emph{\JournalTitle{Phys. Rev. C}}} \textbf{\bibinfo{volume}{63}},
  \bibinfo{pages}{034006} (\bibinfo{year}{2001}).

\bibitem{Herberg:1999ud}
\bibinfo{author}{Herberg, C.} \emph{et~al.}
\newblock \bibinfo{journal}{\bibinfo{title}{{Determination of the neutron
  electric form-factor in the D(e,e' n)p reaction and the influence of nuclear
  binding}}}.
\newblock {\emph{\JournalTitle{Eur. Phys. J.}}} \textbf{\bibinfo{volume}{A5}},
  \bibinfo{pages}{131--135} (\bibinfo{year}{1999}).

\bibitem{Buchmann:2004ia}
\bibinfo{author}{Buchmann, A.~J.}
\newblock \bibinfo{journal}{\bibinfo{title}{{Electromagnetic N ---> $\Delta$
  transition and neutron form-factors}}}.
\newblock {\emph{\JournalTitle{Phys. Rev. Lett.}}}
  \textbf{\bibinfo{volume}{93}}, \bibinfo{pages}{212301}
  (\bibinfo{year}{2004}).

\bibitem{Vanderhaeghen:2007}
\bibinfo{author}{Pascalutsa, V.} \& \bibinfo{author}{Vanderhaeghen, M.}
\newblock \bibinfo{journal}{\bibinfo{title}{{Large-$N_{\rm c}$ relations for the
  electromagnetic nucleon-to-$\Delta$ form factors}}}.
\newblock {\emph{\JournalTitle{Phys. Rev.}}} \textbf{\bibinfo{volume}{D76}},
  \bibinfo{pages}{111501(R)} (\bibinfo{year}{2007}).

\bibitem{Blomberg:2015zma}
\bibinfo{author}{Blomberg, A.} \emph{et~al.}
\newblock \bibinfo{journal}{\bibinfo{title}{{Electroexcitation of the
  $\Delta^{+}(1232)$ at low momentum transfer}}}.
\newblock {\emph{\JournalTitle{Phys. Lett.}}} \textbf{\bibinfo{volume}{B760}},
  \bibinfo{pages}{267--272} (\bibinfo{year}{2016}).

\bibitem{Stave:2006ea}
\bibinfo{author}{Stave, S.} \emph{et~al.}
\newblock \bibinfo{journal}{\bibinfo{title}{{Lowest $Q^2$ Measurement of the
  $\gamma*$ p ---> $\Delta$ Reaction: Probing the Pionic Contribution}}}.
\newblock {\emph{\JournalTitle{Eur. Phys. J.}}} \textbf{\bibinfo{volume}{A30}},
  \bibinfo{pages}{471--476} (\bibinfo{year}{2006}).

\bibitem{Sparveris:2013ena}
\bibinfo{author}{Sparveris, N.} \emph{et~al.}
\newblock \bibinfo{journal}{\bibinfo{title}{{Measurements of the $\gamma$*p
  $\rightarrow$ $\Delta$ reaction at low $Q^{2}$}}}.
\newblock {\emph{\JournalTitle{Eur. Phys. J.}}} \textbf{\bibinfo{volume}{A49}},
  \bibinfo{pages}{136} (\bibinfo{year}{2013}).

\bibitem{Sparveris:2006uk}
\bibinfo{author}{Sparveris, N.~F.} \emph{et~al.}
\newblock \bibinfo{journal}{\bibinfo{title}{{Determination of quadrupole
  strengths in the $\gamma* p \rightarrow \Delta(1232)$ transition at $Q^2$ =
  0.20 $({\rm GeV}/c)^2$}}}.
\newblock {\emph{\JournalTitle{Phys. Lett.}}} \textbf{\bibinfo{volume}{B651}},
  \bibinfo{pages}{102--107} (\bibinfo{year}{2007}).

\bibitem{Blomberg:2019caf}
\bibinfo{author}{Blomberg, A.} \emph{et~al.}
\newblock \bibinfo{journal}{\bibinfo{title}{{Virtual Compton Scattering
  measurements in the nucleon resonance region}}}.
\newblock {\emph{\JournalTitle{Eur. Phys. J. A}}}
  \textbf{\bibinfo{volume}{55}}, \bibinfo{pages}{182} (\bibinfo{year}{2019}).
\newblock \eprint{1901.08951}.

\bibitem{Sparveris:2004jn}
\bibinfo{author}{Sparveris, N.~F.} \emph{et~al.}
\newblock \bibinfo{journal}{\bibinfo{title}{{Investigation of the conjectured
  nucleon deformation at low momentum transfer}}}.
\newblock {\emph{\JournalTitle{Phys. Rev. Lett.}}}
  \textbf{\bibinfo{volume}{94}}, \bibinfo{pages}{022003}
  (\bibinfo{year}{2005}).

\bibitem{Elsner:2005cz}
\bibinfo{author}{Elsner, D.} \emph{et~al.}
\newblock \bibinfo{journal}{\bibinfo{title}{{Measurement of the LT-asymmetry in
  $\pi^{0}$ electroproduction at the energy of the $\Delta(1232)$ resonance}}}.
\newblock {\emph{\JournalTitle{Eur. Phys. J.}}} \textbf{\bibinfo{volume}{A27}},
  \bibinfo{pages}{91--97} (\bibinfo{year}{2006}).

\bibitem{Aznauryan:2009mx}
\bibinfo{author}{Aznauryan, I.~G.} \emph{et~al.}
\newblock \bibinfo{journal}{\bibinfo{title}{{Electroexcitation of nucleon
  resonances from CLAS data on single pion electroproduction}}}.
\newblock {\emph{\JournalTitle{Phys. Rev.}}} \textbf{\bibinfo{volume}{C80}},
  \bibinfo{pages}{055203} (\bibinfo{year}{2009}).

\bibitem{Kelly:2005jy}
\bibinfo{author}{Kelly, J.~J.} \emph{et~al.}
\newblock \bibinfo{journal}{\bibinfo{title}{{Recoil polarization measurements
  for neutral pion electroproduction at $Q^2$ = 1 $({\rm GeV}/c)^2$ near the Delta
  resonance}}}.
\newblock {\emph{\JournalTitle{Phys. Rev.}}} \textbf{\bibinfo{volume}{C75}},
  \bibinfo{pages}{025201} (\bibinfo{year}{2007}).

\bibitem{Alexandrou:2018sjm}
\bibinfo{author}{Alexandrou, C.} \emph{et~al.}
\newblock \bibinfo{journal}{\bibinfo{title}{{Proton and neutron electromagnetic
  form factors from lattice QCD}}}.
\newblock {\emph{\JournalTitle{Phys. Rev.}}} \textbf{\bibinfo{volume}{D100}},
  \bibinfo{pages}{014509} (\bibinfo{year}{2019}).

\bibitem{Ye:2017gyb}
\bibinfo{author}{Ye, Z.}, \bibinfo{author}{Arrington, J.},
  \bibinfo{author}{Hill, R.~J.} \& \bibinfo{author}{Lee, G.}
\newblock \bibinfo{journal}{\bibinfo{title}{{Proton and Neutron Electromagnetic
  Form Factors and Uncertainties}}}.
\newblock {\emph{\JournalTitle{Phys. Lett.}}} \textbf{\bibinfo{volume}{B777}},
  \bibinfo{pages}{8--15} (\bibinfo{year}{2018}).

\bibitem{Xiong:2019umf}
\bibinfo{author}{Xiong, W.} \emph{et~al.}
\newblock \bibinfo{journal}{\bibinfo{title}{{A small proton charge radius from
  an electron–proton scattering experiment}}}.
\newblock {\emph{\JournalTitle{Nature}}} \textbf{\bibinfo{volume}{575}},
  \bibinfo{pages}{147--150} (\bibinfo{year}{2019}).

\bibitem{bonnrn}
\bibinfo{author}{Filin, A.~A.} \emph{et~al.}
\newblock \bibinfo{journal}{\bibinfo{title}{{Extraction of the Neutron Charge
  Radius from a Precision Calculation of the Deuteron Structure Radius}}}.
\newblock {\emph{\JournalTitle{Phys. Rev. Lett.}}}
  \textbf{\bibinfo{volume}{124}}, \bibinfo{pages}{082501}
  (\bibinfo{year}{2020}).

\bibitem{Miller:2007uy}
\bibinfo{author}{Miller, G.~A.}
\newblock \bibinfo{journal}{\bibinfo{title}{{Charge Density of the Neutron and
  Proton }}}.
\newblock {\emph{\JournalTitle{Phys. Rev. Lett.}}}
  \textbf{\bibinfo{volume}{99}}, \bibinfo{pages}{112001}
  (\bibinfo{year}{2007}).

\bibitem{lorce}
\bibinfo{author}{Lorce, C.}
\newblock \bibinfo{journal}{\bibinfo{title}{{Charge Distributions of Moving
  Nucleons}}}.
\newblock {\emph{\JournalTitle{Phys. Rev. Lett.}}} \bibinfo{pages}{in press,
  arXiv:2007.05318} (\bibinfo{year}{2020}).


\end{thebibliography}

\end{document}